\documentclass[letter,11pt]{article}
\usepackage{float}
\usepackage{graphics,graphicx}
\usepackage{helvet}
\usepackage[utf8]{inputenc}
\usepackage[T1]{fontenc}
\usepackage{longtable}
\usepackage{soul}
\usepackage{amsmath, amssymb, gensymb}
\usepackage{natbib}
\usepackage{microtype}
\usepackage{multirow}
\usepackage{morefloats}
\usepackage{lscape}
\usepackage{multicol}
\usepackage{aas_macros}
\usepackage{hyperref,setspace}

\def\eg {e.g.,} %e.g.,
\def\kms {km\,s$^{-1}$}

\def\mjybm {mJy\,bm$^{-1}$}
\def\ujybm {$\mu$Jy\,bm$^{-1}$}
\def\njybm {nJy\,bm$^{-1}$}

\newcommand\arcsec{\mbox{$^{\prime\prime}$}}%

\begin{document}

\begin{center}
\includegraphics[width=\textwidth]{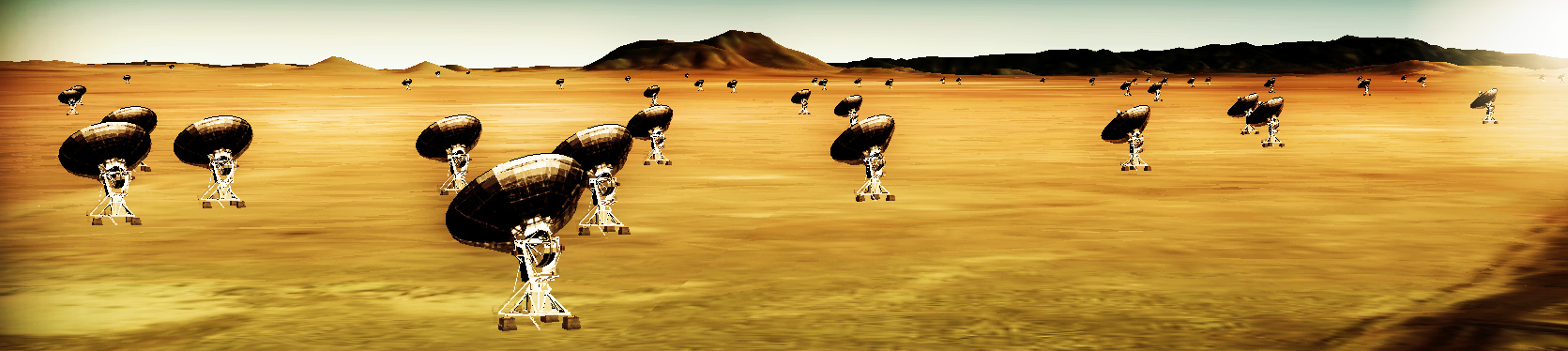}
\end{center}

\begin{center}

{\bf \Large Next Generation Very Large Array Memo No. 6}

\vspace{0.1in}

{\bf \Large Science Working Group 1}

\vspace{0.1in}

{\bf \Large The Cradle of Life}

\end{center}

\hrule

\vspace{0.7cm}

%\noindent Cradle of Life Working Group

\noindent Andrea Isella$^1$, Charles L. H. Hull$^{2,3}$, Arielle Moullet$^4$, Roberto Galv\'an-Madrid$^5$, Doug Johnstone$^6$, Luca Ricci$^2$, John Tobin$^8$,
Leonardo Testi$^7$, Maite Beltran$^{22}$, Joseph Lazio$^{9}$, Andrew Siemion$^{10,23,24}$, Hauyu Baobab Liu$^{11}$, Fujun Du$^{16}$,
Karin I. \"Oberg$^2$, Ted Bergin$^{16}$, Paola Caselli$^{14,15}$, Tyler Bourke$^{17}$,  Chris Carilli$^{12,25}$, Laura Perez$^{12}$, Bryan Butler$^{12}$, Imke de Pater$^{10}$, Chunhua Qi$^2$, Mark Hofstadter$^{9}$,
Raphael Moreno$^{13}$, David Alexander$^1$, Jonathan Williams$^{18}$, Paul Goldsmith$^9$, Mark Wyatt$^{19}$, Laurent Loinard$^5$, James Di Francesco$^{6}$,
David Wilner$^2$, Peter Schilke$^{20}$, Adam Ginsburg$^7$, \'Alvaro S\'anchez-Monge$^{20}$, Qizhou Zhang$^2$, Henrik Beuther$^{21}$ \\

\vspace{0.7cm}

\begin{center}
{\bf \large Abstract}\\
\end{center}

This paper discusses compelling science cases for a future
long-baseline interferometer operating at millimeter and centimeter
wavelengths, like the proposed \textit{Next Generation Vary Large Array}
(ngVLA). We report on the activities of the \textit{Cradle of Life}
science working group, which focused on the formation of low- and
high-mass stars, the formation of planets and evolution of
protoplanetary disks, the physical and compositional study of Solar
System bodies, and the possible detection of radio signals from
extraterrestrial civilizations. We propose 19 scientific projects
based on the current specification of the ngVLA. Five of them are
highlighted as possible Key Science Projects: (1) Resolving the
density structure and dynamics of the youngest HII regions and
high-mass protostellar jets, (2) Unveiling binary/multiple protostars
at higher resolution, (3) Mapping planet formation regions in nearby
disks on scales down to 1 AU, (4) Studying the formation of complex
molecules, and (5) Deep atmospheric mapping of giant planets in the
Solar System.  For each of these projects, we discuss the scientific
importance and feasibility.  The results presented here should be
considered as the beginning of a more in-depth analysis of the science
enabled by such a facility, and are by no means complete or
exhaustive.

\vspace{6mm}

\begin{spacing}{1}
\footnotesize
\noindent $^{1}$ Physics \& Astronomy Department, 6100 Main MS-61, Houston, Texas 77005, USA, \texttt{isella@rice.edu}\\
\noindent $^{2}$ Harvard-Smithsonian Center for Astrophysics, 60 Garden St., Cambridge, MA, 02138, USA, \texttt{chat.hull@cfa.harvard.edu}\\
\noindent $^{3}$ Jansky Fellow of the National Radio Astronomy Observatory\\
\noindent $^{4}$ National Radio Astronomy Observatory, 520 Edgemont Road, Charlottesville, Virginia 22903, USA, \texttt{amoullet@nrao.edu}\\
\noindent $^{5}$ Instituto de Radiostronom\'ia y Astrof\'isica, Universidad Nacional Aut\'onoma de M\'exico, 58089 Morelia, Michoac\'an, M\'exico\\
\noindent $^{6}$ National Research Council of Canada, Herzberg Institute of Astrophysics, 5071 West Saanich Road, Victoria, BC V9E 2E7, Canada\\
\noindent $^{7}$ ESO, Karl Schwarzschild str. 2, 85748 Garching, Germany\\
\noindent $^{8}$ Leiden Observatory, Leiden University, P.O. Box 9513, 2300-RA Leiden, The Netherlands \\
\noindent $^{9}$ Jet Propulsion Laboratory, California Institute of Technology, 4800 Oak Grove Dr., M/S 67-201, Pasadena, CA 91109, USA \\
\noindent $^{10}$ Department of Astronomy, 501 Campbell Hall, University of California, Berkeley CA 94720\\
\noindent $^{11}$ Academia Sinica Institute of Astronomy and Astrophysics, P.O. Box 23-141, Taipei, 106, Taiwan\\
\noindent $^{12}$ National Radio Astronomy Observatory, 1003 Lopezville Road, Socorro, New Mexico, 87801, USA\\
\noindent $^{13}$ LESIA, Observatoire de Paris, 5 place J. Janssen, Meudon 92190, France\\
\noindent $^{14}$ Max Planck Institute for Extraterrestrial Physics, Giessenbachstrasse 1, D-85748 Garching, Germany\\
\noindent $^{15}$ School of Physics and Astronomy, University of Leeds, Leeds LS2 9JT, UK\\
\noindent $^{16}$ Department of Astronomy, University of Michigan, 1085 S. University Ave., Ann Arbor, Michigan 48109, USA\\
\noindent $^{17}$ SKA Organisation, Jodrell Bank Observatory, Lower Withington, Macclesfield, Cheshire SK11 9DL, UK\\
\noindent $^{18}$ Institute for Astronomy, University of Hawaii, 2680 Woodlawn Dr., Honolulu, HI 96822, USA\\
\noindent $^{19}$ Institute of Astronomy, Madingley Rd, Cambridge CB3 0HA, UK\\
\noindent $^{20}$ Physikalisches Institut, Universit\"at zu K\"oln, Z\"ulpicher Str. 77, D-50937 K\"oln, Germany \\
\noindent $^{21}$ Max Planck Institute for Astronomy, K\"onigstuhl 17, 69117 Heidelberg, Germany \\
\noindent $^{22}$ INAF-Osservatorio Astrofisico di Arcetri, Largo E. Fermi 5, I-50125 Firenze, Italy \\
\noindent $^{23}$ Department of Astrophysics, Radboud University, Heyendaalseweg 135, 6525 AJ Nijmegen, Netherlands \\
\noindent $^{24}$ ASTRON, Netherlands Institute for Radio Astronomy, PO Box 2, 7990 AA Dwingeloo, Netherlands \\
\noindent $^{25}$  Cavendish Astrophysics Group, Cambridge, UK \\
\end{spacing}

\newpage

\tableofcontents

% Introduction
\clearpage
\section{Executive Summary}

\begin{figure} [hbt!]
\centering
\includegraphics[scale=0.5, clip, trim=0cm 0cm 0cm 0cm]{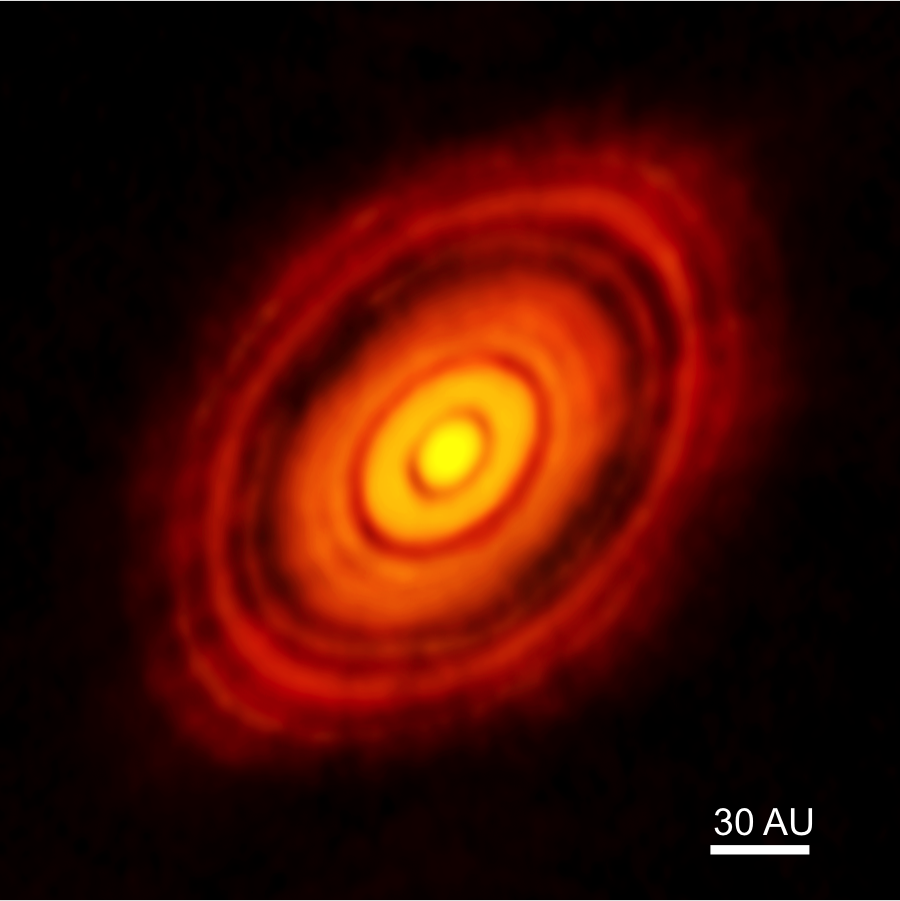}
\caption{\footnotesize   
ALMA image of HL Tau, a protostar with an embedded disk that shows signs of planet formation.  The region inside the first gap is optically thick ($T > 200$\,K); additional outer rings may be optically thick as well.  This image is based on Figure 2 of \citet{hltau}.  Image source: \texttt{almaobservatory.org}.
}
\label{ch1_fig:HLTau} 
\end{figure}

The study of the formation of stellar systems---from protostars, to protoplanetary disks, down to fully evolved planetary systems---is one of the fastest growing research fields in astrophysics. It has benefitted tremendously in the past decade from the synergy of cutting edge instruments covering the thermal domain (Herschel, JVLA, SMA, CARMA, IRAM), where planet-forming dust, molecules and plasmas are all most readily observed. For example, the international ALMA project has opened exceptional pathways for the study of protoplanetary disks, as demonstrated by the recent observations of the disk around the low mass star HL Tau (\citealt{hltau}; see Figure \ref{ch1_fig:HLTau}). In the future, progress in the field will be contingent upon the availability of comparably high spatial resolution at all thermal wavelengths.

This white paper presents a series of specific science cases that would be made possible by what is currently designated the ``Next Generation Very Large Array'' (ngVLA), a cutting-edge interferometer operating from 1--115\,GHz with ten times the effective collecting area of the JVLA and ALMA.  The ngVLA would have ten times longer baselines (300\,km), providing milli-arcsecond resolution, and would also have a dense core of antennas on kilometer scales for high surface brightness imaging \citep{Carilli2015}.  The exquisite spatial resolution and sensitivity of the ngVLA will enable major breakthroughs in the fields of star formation, planet formation, and planetary science.

In particular, we identify the following five {\it Key Science Projects} (KSPs):\footnote{\,We have chosen to present the KSPs in order of decreasing distance to the prime targets; this is not an implication of priority.}

\begin{description}
\item \textbf{Resolving the density structure and dynamics of the youngest HII regions and high-mass protostellar jets (\S\,\ref{ch2_KSP})}

While improvements in observations, numerical models, and theory have led to a reasonable understanding of the formation of nearby individual or binary stars---which are almost exclusively low-mass---our understanding of the formation of massive stars and their natal clusters is still quite uncertain. This is largely due to the inherent difficulty in obtaining observations of massive star-forming regions, both because the sources are located further from the Sun and because the environments are much more complex \citep[for a recent review see][]{Tan14}. The clearest signposts for massive stars are their surrounding HII regions, which have been the subject of much theoretical work \citep[e.g.,][]{Peters10,Keto03,Franco00}. The ngVLA, with its superior spatial resolution and ability to detect the centimeter-wavelength free-free emission that originates in these ionized environs, will play a key role in unravelling the mystery of high mass star formation and evolution.

\item \textbf{Unveiling binary/multiple protostars at higher resolution and in more distant star-forming regions (\S\,\ref{ch3_KSP})}

A key question in star formation is how a collapsing core fragments to form binary, or multiple, stellar systems. 
Present (sub)millimeter interferometers have the resolving power to peer deep inside the enshrouding envelope, but they are 
 hampered by the high dust opacity that characterizes these regions. As a result of the high optical depth, pre-stellar cores 
do not appear to have significant substructure, 
and yet by the time the protostars become visible through the envelope, binary and multiple systems have become quite common. 
Centimeter-wavelength observations with the ngVLA will suffer from much lower opacity, allowing the earliest possible detection of 
binarity on AU scales within star-forming cores.  

%The improved statistics of protostellar multiplicity will help to elucidate the conditions 
%that lead to the current observable state of multiplicity in main sequence stars: i.e., that nearly all high mass (O, B) stars are in 
%multiple systems, whereas only about 50\% of FGK field stars, $\sim$\,33\% of M stars, and <\,30\% of L and T dwarfs are in multiples \citep[e.g.,][]{DucheneKraus2013}.

\item \textbf{Mapping planet formation regions in nearby disks (\S\,\ref{ch4})}

In the earliest stages of their formation, planets are enshrouded by dense circumstellar disks, whose thermal emission is optically 
thick even at millimeter wavelengths. By operating at centimeter wavelengths, the ngVLA will allows us to map planet formation regions down 
to innermost few AU of these disks.  Furthermore, observations over a range of wavelengths will allow for the removal of any free-free emission 
caused by jets or winds emanating from the central protostar, thus enabling the determination of the thermal emission spectral index---a robust and key 
measure of the grain growth that leads to planetary formation. Another poorly constrained characteristic of disks is their temperature structure. The ngVLA will provide access to the most commonly used thermometer for interstellar and circumstellar media---NH$_3$---at sufficient sensitivity to measure and spatially resolve the temperature structure of nearby disks, thus benchmarking theories of planet formation. NH$_3$ observations are also important to map out the nitrogen reservoirs in disks: nitrogen is one of the key ingredients of the molecules of life, but its distribution and abundance during planet formation is unknown due to a lack of observational constraints on the predicted main carriers.

\item \textbf{Studying the formation of complex molecules. (\S \ref{ch4.5})}

Centimeter astronomy presents unique access to transitions from very complex organic 
molecules that are ``crowded out'' at shorter (ALMA) wavelengths by very high densities 
of lines emitted from smaller molecules. The elusive and prebiotically important glycine may only 
be possible to detect with the ngVLA. Similarly, the ngVLA could provide the first interstellar detections of sugars, 
which are the starting point of RNA and DNA backbones.

\item \textbf{Deep atmospheric mapping of giant planets in the Solar System (\S\,6})

Within our Solar System centimeter-wavelength ngVLA observations will allow for key observations of the subsurfaces of rocky bodies and of the denser, deeper levels of giant planet atmospheres. The latter observations are necessary to disentangle the vertical temperature profile, the composition, and the distribution of molecular absorbers. The enhanced sensitivity of the ngVLA will also allow for integration times significantly shorter than the planet rotation period, providing good temporal mapping resolution for these planets.

\end{description}

\bigskip
In the sections that follow we describe the main scientific goals of the KSPs and the other science cases.

% High-mass star formation
\clearpage
\section{Formation of high-mass stars}
\label{ch2}

\subsection{Resolving the density structure and dynamics of the youngest HII regions and high-mass protostellar jets}
\label{ch2_KSP}

The JVLA has spatially resolved and mapped the dynamics (via hydrogen recombination lines, or RLs) 
of the ``ultracompact'' (UC) and some ``hypercompact'' (HC) HII regions around groups of young O-type stars 
\citep[see the reviews by][]{Hoare07,Churchwell02}. 
Roughly, free-free emitting objects at a distance of about 5\,kpc have flux densities $S_{1.3\,\textrm{cm} } \sim$ 0.1--1\,Jy,
optical depths $\tau_{1.3\,\textrm{cm}} \sim$\,0.1--1, and spatial sizes between  0.03--0.3\,pc. 

\begin{figure} [hbt!]
\centering
\includegraphics[scale=0.6, clip, trim=1cm 0cm 0cm 0cm]{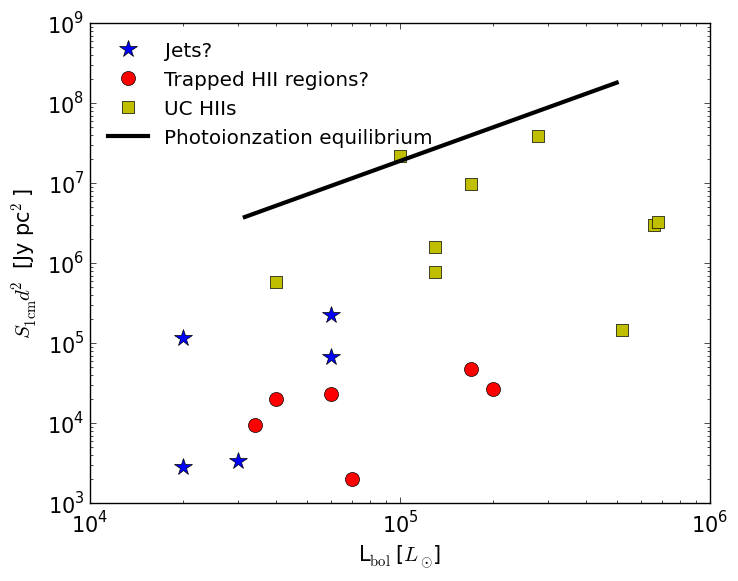}

\caption{\footnotesize   
Ultracompact HII regions have radio fluxes close to the expectation from photoionization equilibrium and 
most likely are young, massive 
($M_\star > 20~M_\odot$) stars with no further accretion. However, there is a population of fainter ionized sources that current 
facilities like the JVLA can only detect but cannot resolve. These sources could be small radio jets, stellar winds, or HII regions that are gravitationally trapped. 
Determining their nature is key for understanding the final stages of accretion in the formation of massive stars.
} 
\label{ch2_fig:faintradio} 
\end{figure}

However, the faintest objects---those that may be in transition from a high-mass (shock ionized) protostellar jet 
to a (photo ionized) HII region---are currently detected only in continuum, and are not resolved 
\citep[e.g.,][]{Rosero14,vanderTak2005}. These objects (see Figure~\ref{ch2_fig:faintradio}) are
presumably in the latest phase of accretion in the formation of stars more massive than 10 or 20 M$_\odot$
and might have characteristic radii of 100--200 AU, corresponding to the gravitational radius of 10--20 M$_\odot$
stars \citep{Lizano08, Keto2002}.  
An angular resolution between 20--40 mas is therefore required to resolve these objects at the distance of 5 kpc. 
With an expected angular resolution of 6--12 mas at 50 GHz, the ngVLA will therefore resolve with several beams 
the smallest HII regions. 

The smallest photoionized HII regions are known to be variable over periods of months to years \citep[e.g.,][]{DePree14,GM08}. 
Radio jets can be 
variable over timescales as short as days \citep{Liu14}. Simultaneous observations at multiple frequencies are therefore key for 
deriving the density structure. This requirement is met by the preliminary ngVLA design, which delivers 
simultaneous observations between 5--50 GHz, with a stretch goal of 1--100 GHz.

Finally, the stellar winds from young stars already on the main sequence could be routinely detected with the ngVLA. 
Currently, the JVLA can only detect the thermal emission of stellar winds from stars more massive than approximately O9. 
The ngVLA could directly detect the stellar winds of B-type stars, expected to be at the level of  a few $\mu$Jy  
\citep[][]{Dzib13} in less than one hour of integration.

\bigskip
\textit{Frequency requirement:} 10--50\,GHz for radio continuum.  For recombination lines, the range between 40--100 GHz is ideal because RLs are brighter than at lower frequencies \citep[e.g.,][]{Guzman14,KZK08}, but are not yet contaminated by molecular lines as is often the case in the ALMA range at $> 100$ GHz. 

\bigskip
\textit{Sensitivity requirement}: the brightness of resolved, faint HII regions at 50, 25, 12, and 8 GHz will be 22, 50, 62, and 145\,$\mu$Jy beam$^{-1}$, respectively, to be compared with the 0.3 $\mu$Jy beam$^{-1}$ noise level achievable with the ngVLA in 1 hr of integration of source. 
For recombination lines, considering the factor $\times10$ improvement in sensitivity with respect to the JVLA, 
individual recombination lines (RLs) can be mapped in the youngest HII regions at 45\,GHz with 20 mas resolution 
with integration times of tens of hours. However, RLs with similar quantum numbers can be stacked together to increase the signal to noise or to decrease 
the integration time. 

\bigskip
\textit{Angular resolution requirement:}
Assuming an angular size of 40 mas, the ngVLA will resolve the emission from 
the smallest HII regions with tens of resolution elements across at frequencies above 25 GHz.

\subsection{ \sloppy{High resolution centimeter-wavelength observations of molecular line absorption against background ionized sources} }

\begin{figure} [hbt!]
\centering
\includegraphics[scale=0.45, clip, trim=0cm 0cm 0cm 0cm]{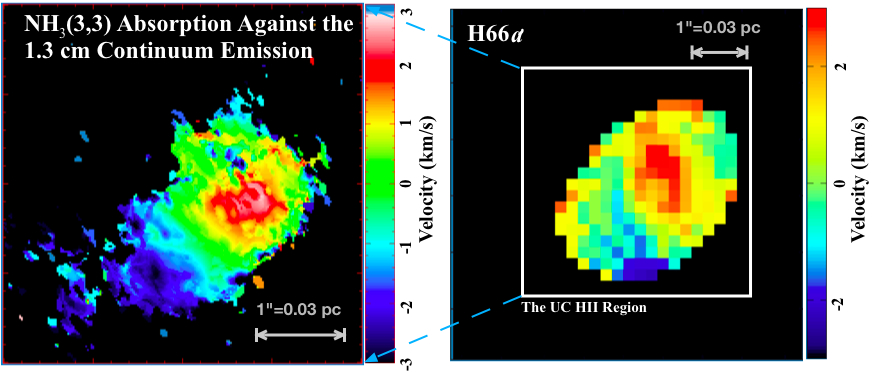}
\caption{\footnotesize   
JVLA observations of the molecular (left) and ionized (right) accretion flows in G10.6--0.4.  The left panel shows the first moment maps 
of NH$_3$ absorption against the free-free background of the UCHII region; the right panel shows the H66$\alpha$ RL in emission. 
In this case the JVLA could resolve the accretion flows because the stellar mass inferred from the kinematics is as large as a 
few hundred $M_\odot$ (most likely an aggregate of massive stars). The ngVLA is needed to resolve the molecular and ionized accretion 
flows around individual or binary massive stars. Reproduced from \cite{Sollins05}. 
} 
\label{ch2_fig:ionziedacc} 
\end{figure}

Molecular absorption against a bright (high T$_B$) background continuum source is possibly the clearest way of diagnosing infall 
and accretion \citep[e.g.,][]{GM09,Beltran06,HoHaschick86}. The JVLA has unambiguously diagnosed 
infall and accretion via this technique toward bright ultracompact HII regions down to 100 mas angular resolution
\citep[e.g.,][see Figure~\ref{ch2_fig:ionziedacc}]{Liu11,Sollins05}. The ngVLA can make these diagnostics around fainter ionized 
sources at $\times10$ better angular resolution. For example, direct accretion could 
be detected toward embedded low-mass stars, or toward massive stars just after the onset of their HII region. 
The ngVLA would also enable time-domain accretion studies by looking at variability in the molecular absorption.

\bigskip
\textit{Frequency requirement:} the obvious choices are NH$_3$ lines between 18--40\,GHz, but other lines are possible. 

\bigskip
\textit{Sensitivity requirement:} embedded HII regions are expected to have brightness temperatures of several thousand K. 
Achieving an rms brightness temperature of $\sim$100 K in 1 km/s velocity channels is ideal for this purpose. %Lines can be stacked too. \\

\bigskip
\textit{Angular resolution requirement:} $\sim$20 mas resolution is sufficient to resolve the accretion flows surrounding the nearest hypercompact HII regions (and other ionized sources) at <100 AU scales. Field of view is not an issue since these sources are typically only a few arcsec wide.

\subsection{Linearly polarized dust emission and synchrotron polarization in OB star- or cluster-forming cores}

The ngVLA would allow us to understand the role of magnetic fields in the formation of massive stars and clusters, 
from 20,000 AU (0.1 pc, or 3\arcsec\ at 6 kpc) to 100 AU scales. The ngVLA could map the polarization in massive star formation regions, where magnetic fields may be important for halting fragmentation and may 
set star formation efficiencies. Polarization related to synchrotron could be detected in the centimeter bands 
\citep[e.g.,][]{CG10}, whereas 
dust polarization could be observed in the millimeter bands \citep[e.g.,][]{Zhang14,Girart06}. The ngVLA long millimeter 
bands have clear advantages over 
shorter-wavelength interferometers, since opacities are low in the denser, smaller-scale regions.

\bigskip
\textit{Frequency requirement:} observations at 40--50 GHz are particularly useful because many objects are line-crowded in the millimeter 
band (e.g., Orion BN-KL), which prohibits continuum/line separation. In addition, on $\sim$100 AU scales, these regions are optically 
thick in short millimeter band observations. Finally, long-millimeter wavelengths have the advantage that de-polarization due to scattering is negligible.  

\bigskip
\textit{Sensitivity requirement:} it is difficult to predict the strength of the polarized emission. Synchrotron polarization 
has been imaged with the JVLA in a couple of landmark objects \citep{PerezSanchez13,CG10}. 
Linearly polarized dust emission has been detected at millimeter wavelengths in protostellar 
envelopes \citep[e.g.,][]{Girart06, Hull2014}, as well as in a few disks \citep[e.g.,][]{Stephens14}.

\subsection{Star-formation in the inner few pc of the Galactic center}

The JVLA has mapped the ionized mini-spirals within the circumnuclear disk (CND) within 5 pc of the Galactic 
Center, Sgr A* \citep[e.g.,][]{Zhao09}. 
However, only with the sensitivity and angular resolution of the ngVLA will we be able to detect the young 
stellar object (YSO) population within the CND and beyond, to address the long-standing question of whether star formation 
around the Galactic Center is possible, and if so, how does it proceed \citep[for a review, see][]{Genzel10}. 
The ngVLA will be able to detect and resolve the entire putative YSO population, low- and high-mass. 
The ngVLA could also measure the orbital motions (proper motions and line-of-sight) of the star-forming gas reservoirs within 
and beyond the CND \citep[e.g.,][]{Liu12}---such as peculiar objects like the G2 cloud \citep{Gillessen12}---to 
test scenarios of stellar migration and address how the stellar 
disks around Sgr A* are formed.

\bigskip
\textit{Frequency requirement:} 27--50\,GHz is ideal for avoiding the confusion either from ambient dust emission, or from the 
ionized mini-spiral arms. The spectral slope from 27\,GHz to > 50 GHz bands can help diagnose the thermal emission mechanism. 
Since the Galactic Center is the most obscured SF region in the Milky Way, centimeter-band observations of dust emission are highly 
preferred. 
The simultaneous 3:1 frequency coverage would greatly aid this science case, since the different emission mechanisms 
of the possible YSOs could be imaged at the same time: synchrotron from stellar magnetospheres, free-free from radio jets 
or stellar winds, and dust from circumstellar disks and envelopes. 

\bigskip
\textit{Sensitivity requirement:} the estimates depend on the assumption of the dust opacity. Assuming a radio dust opacity slope $\beta=1$ or smaller, 
the ngVLA continuum noise level of 100\,\njybm{} (10 hrs on source) can detect objects that have masses comparable to  
the Classical T Tauri disks in nearby star-forming regions. Combining with kinematic information from molecular line observations,
ngVLA maps  can trace 3D orbits of individual (proto)stellar objects or gas cores. 

%(does not need very good resolution for line observations; maybe can 
%complement the line part with ALMA) 

\bigskip
\textit{Angular resolution requirement:} 10--20 mas angular resolution, to provide 80--160 AU physical resolution, is adequate to 
resolve individual (proto)stellar accretion disks (if there are any). High resolution observations are required to filter out the 
extended structure.  10--20 mas resolution is adequate for the purpose of observing proper motions, in an analogous way 
to what has been done with stellar objects using telescopes such as the VLT and Keck.
\citep[e.g.,][]{Gillessen09,Ghez08}. 

\bigskip
\textit{Field of view requirement:} The CND is 1--2$\arcsec$ wide, and can be mapped with 25 m dishes by performing a small mosaic. 
Large dishes are better than small dishes since this would make it easier to calibrate the long baseline data at high frequency.

% Low-mass star formation
\clearpage

\section{Formation of low-mass stars}
\label{ch3}

Probing the inner regions of the disks and cores around very young protostars is critical to understanding 
the earliest stages of disk, outflow, jet, and binary/multiple formation. This requires high-angular resolution 
observations at centimeter wavelengths in order to map both the spatial distribution of dust and gas as well as 
the kinematics in the densest regions, which are opaque at high at millimeter wavelengths.  
The low frequency, high resolution, and excellent sensitivity of the ngVLA will open 
a new observational window, allowing us to study 
the physical processes that control the formation of low mass stars in the innermost 
dense regions that are unreachable by current telescopes.   
The main avenues we discuss in this section are the study of binary/multiple stars,
and the structure of the densest regions of protostellar cores, disks, and jets.

%Other areas relevant to both younger Class~0/I sources and more evolved Class II objects include mapping the planet-forming regions
%in nearby disks (see Section \ref{ch4_KSP}), 
%measuring the radial distribution of dust grain sizes (see Section \ref{ch4_sec:dust}),
%and probing the cold molecular chemistry of planet-forming regions (see Section \ref{ch4_sec:chemistry}).

\subsection{Binary/multiple stars}
\label{ch3_KSP}

\begin{figure} [hbt!]
\centering
\includegraphics[scale=0.2, clip, trim=1cm 0cm 0cm 1cm]{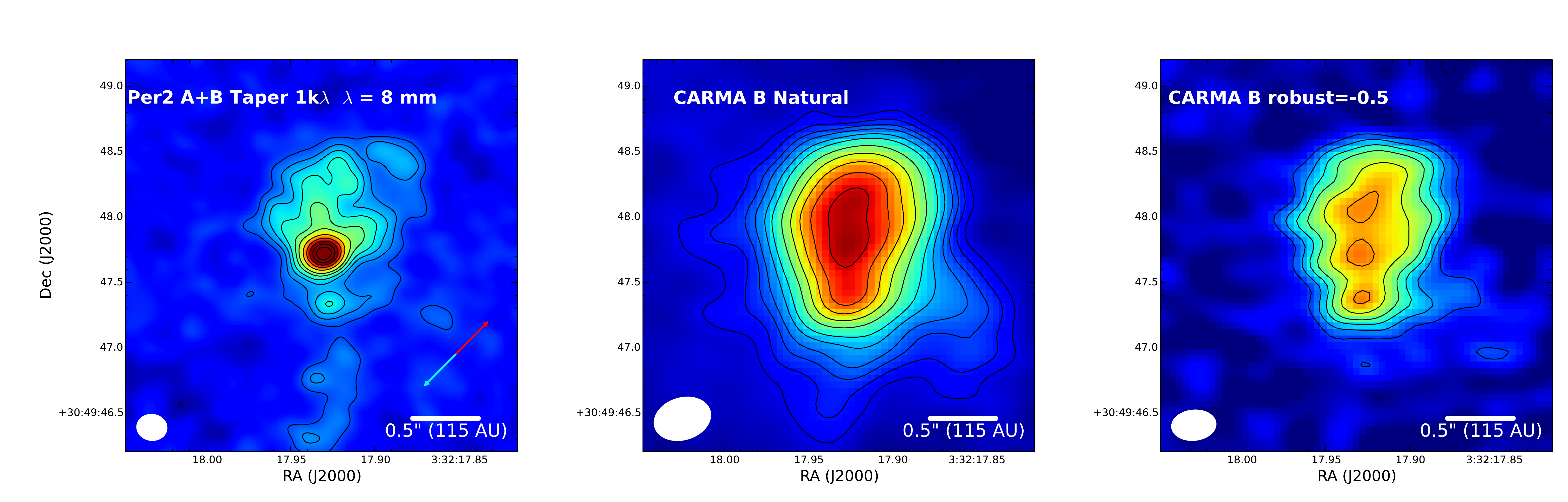}
\caption{\footnotesize   
A comparison of observations toward IRAS 03292/Per2: the VLA 8\,mm data (left, Tobin et al. 2015, in prep.) and CARMA 1.3\,mm data (right, \citealt{Tobin2015b}) have nearly the same angular resolution. The 1.3\,mm data appear to be optically thick, given the smooth spatial distribution at the highest surface brightness level. This source may have a self gravitating disk---there is evidence for rotation on $\sim$\,1000\,AU scales, and the mass of the strucuture is 0.1--0.2\,M$_{\odot}$.  The rms noise values are, from left to right, 11.3\,\ujybm{} (8\,mm), 1.15\,\mjybm{} (1.3\,mm), and 1.71\,\mjybm{} (1.3\,mm).  
} 
\label{ch2_fig:disks} 
\end{figure}

\begin{figure} [hbt!]
\centering
\includegraphics[scale=0.3, clip, trim=0cm 0cm 0cm 1.2cm]{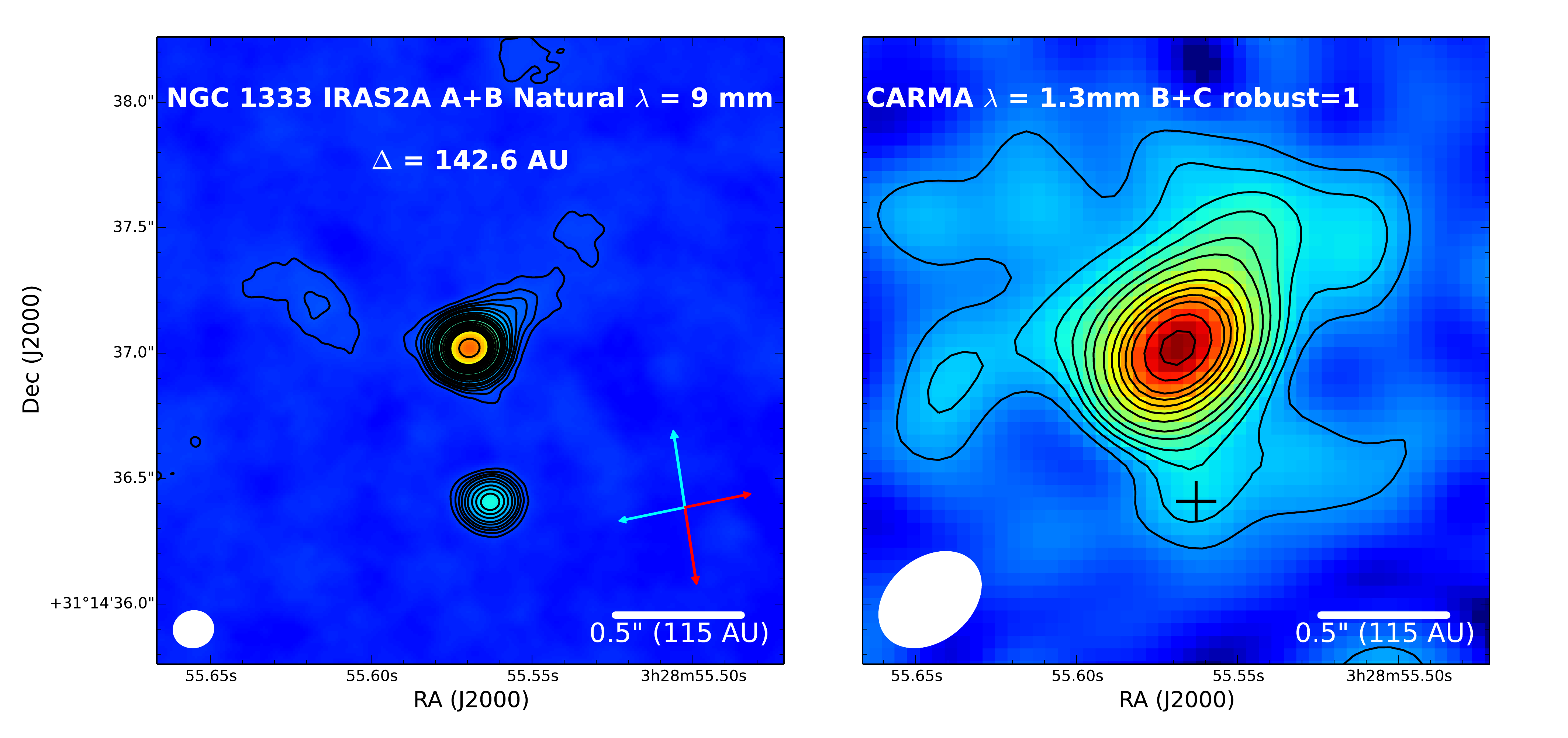}

\caption{\footnotesize
Maps of NGC 1333 IRAS 2A  obtained with the JVLA at the wavelength of 9\,mm (left) and with CARMA 
at the wavelength of 1.3~mm (right) (\citealt{Tobin2015a}). The higher angular resolution of JVLA observations, 
combined with the lower optical depth of the 9~mm emission, resolves two components of IRAS 2A that are separated 
by about 0.5\arcsec.  The rms noise values in the IRAS2A images are 5.4\,\ujybm{} (9mm), and 1.5\,\mjybm{} (1.3mm).
Contours are plotted every 5$\times$ the noise level. 
%1.3 mm contours are plotted at 3, 5, 7, 9, 12, 15, 20, 25, 30, 35, 40, 50, ... $\times$ the rms noise, while in all images except for the 9\,mm IRAS 2A image (bottom left), %which has contours plotted at 5, 10, 15, 20, 25, 30, 35, 40, 50, ... $\times$ the rms noise.
} 
\label{ch2_fig:multiplicity} 
\end{figure}

The innermost regions of molecular cores and newborn protostellar disks are expected to 
be optically thick even at millimeter wavelengths, hampering the ability to study the formation of 
multiple systems with ALMA (see Table \ref{table:optical_depth} and Figures \ref{ch2_fig:disks} and \ref{ch2_fig:multiplicity}; \citealt{Tobin2015a}).
However, high angular resolution observations at wavelengths longer than 3~mm will allow us to probe very close 
binaries that are still embedded in the parent envelopes or disks. For example, at 1 cm, an angular resolution of 
0.01\arcsec\ will enable us to resolve binaries with $\sim$5\,AU separation at distances out to $\sim$\,500\,pc, allowing us 
to collect much better statistics of protostellar multiplicity in nearby star-forming regions. 

While ALMA generally probes the thermal dust and molecular emission from protostars, emission from 
ionized gas (e.g., free-free emission) is very important for seeing embedded sources, and complements millimeter observations
\citep{Anglada1998, Rodriguez1998, Reipurth2002, Reipurth2004, Tobin2015a}.  
By using the ngVLA we will see the free-free component, which is another energetic diagnostic inside a collapsing envelope.

Finally, it is key to note that by resolving close separation ($< 10$ AU) binary systems, long-baseline cm wave observations will 
enable binary motion to be observed on time scales of few years or even months, therefore providing a tool to directly measure  
the mass of protostars. 

\begin{table}[hbt!]
\centerline{Table \ref{table:optical_depth}: disk optical depth as a function of wavelength \vspace*{0.1in}}
\begin{center}
{\normalsize
\begin{tabular}{lr}
\hline
Frequency & Radius \\
                  & (AU) \\
\hline
450\,$\mu$m & 73.4$^\dagger$ \\
850\,$\mu$m & 38.8\phantom{$^\dagger$} \\
1.3\,mm & 25.4\phantom{$^\dagger$} \\
3\,mm & 11.0\phantom{$^\dagger$} \\
8\,mm & 4.2\phantom{$^\dagger$} \\
1\,cm & 3.3\phantom{$^\dagger$} \\
3\,cm & 1.2\phantom{$^\dagger$} \\
\hline
\end{tabular}
\bigskip

$^\dagger$\,In this case the entire disk is optically thick.

\caption{\footnotesize 
Radii as a function of wavelength interior to which a typical circumstellar disk becomes optically thick.  This assumes a disk with $r = 50$\,AU, $M = 0.1$\,$M_{\odot}$, and surface density $\propto$\,$r^{-1}$. Further assumptions: $\kappa_\textrm{1.3\,mm} = 0.9$\,cm$^2$/g (dust only); $M_{\rm disk} = M_{\rm gas} + M_{\rm dust}$, with a dust-to-gas ratio of 1:100; dust opacity scales as $\beta = 1$. Note that this is for face-on disks; with any inclination, optical-depth effects will be worse.
}
\label{table:optical_depth}
}
\end{center}
\end{table}

%\begin{figure} [hbt!]
%\centering
%\includegraphics[scale=0.5, clip, trim=0cm 0cm 0cm 1.2cm]{tobin_optical_depth}
%\caption{\footnotesize   
%Visibility amplitude profiles for disks with a fixed mass, but varying radius at 1.3\,mm (right) and 8.1\,mm (left).  The visibility amplitude profiles show the overall difference in flux density depending on the size of the disk, attributable directly to optical depth. But also the profiles differ considerable at uv-distances longer than 500 k$\lambda$.  This can also be attributed to optical depth: once the resolution becomes equivalent to the size of the optically thick region in the disk, the region will be resolved out because the intensity profile becomes roughly constant.  These models were run with the Hyperion radiative transfer code \citep{Robitaille2011}, are inclined by 60$\degree$, and have a surface density profile proportional to r$^{-1}$. The dust opacities assumed in this model are from \citet{Ossenkopf1994} Table 1, Column 5; $\beta = 1.78$ for this dust model. No envelope is included.  See \citet{Tobin2015b}; note that the plots shown here have been extended to longer $uv$-distances than those in the paper.
%} 
%\label{ch2_fig:optical_depth} 
%\end{figure}

\bigskip
\textit{Frequency requirement:}
Dust and molecular thermal and free-free emission can be used to probe multiplicity in protostars at high resolution. 
the optimal frequency range is $\sim$\,10--30\,GHz. Observations at higher frequencies will be hampered by the high optical depth 
of the innermost regions of molecular cores and protostellar disks, while observations at lower frequency will suffer from 
poorer angular resolution and signal-to-noise ratio. 

\bigskip
\textit{Sensitivity requirement:}
The environment surrounding protobinary systems is expected to be warm, with temperatures 
ranging from $\sim$\,100--1000 K.  A sensitivity between 1--10 K in the continuum should therefore 
be sufficient to obtain high signal-to-noise ratio maps of these systems.  At 30 GHz, and assuming 
a beam size of 0.01\arcsec\, a sensitivity of 1 K corresponds to a rms noise level of about 70 nJy. 
 
\bigskip
\textit{Angular resolution requirement:}
At 30 GHz, an angular resolution of 0.01\arcsec\ is required to resolve protobinary systems with a 
separation of 5 AU at the distance of 500 pc.  This resolution will allow us to resolve a protobinary system with 5 AU separation protobinary system 
in Orion, and a system with 1 AU separation in the Taurus and Ophiuchus star forming regions ($d = 130$\,pc). 
At 10 GHz, an angular resolution of 0.03\arcsec\ will enable us to resolve the youngest protobinary systems, which are 
still embedded in a very dense environment, down to a separation of 4 AU  and 15 AU, at the distances of 130 pc and 500 pc, 
respectively.

\subsection{Dust polarization in low-mass protostellar objects}

Mapping the morphology of magnetic fields in the cores and envelopes surrounding young, low-mass protostars is critical to further our understanding of how magnetic fields affect the star-formation process at early times.  Under most circumstances, spinning dust grains are thought to align themselves with their long axes perpendicular to magnetic field lines \citep[\eg{}][]{Hildebrand1988, Lazarian2003, Lazarian2007, Hoang2009, Andersson2012}, and thus the radiation from the grains is polarized perpendicular to the magnetic field.  Thus, this type of observation can be used to infer the magnetic field morphology in the plane of the sky.

Polarized thermal dust emission from both low- and high-mass protostellar objects has been studied extensively at millimeter wavelengths and at high resolution (1--2$\arcsec$) by the SMA \citep[e.g.,][]{Girart06, ZhangQ2014} and CARMA \citep[e.g.,][]{Hull2013, Hughes2013, Hull2014, Stephens14, SeguraCox2015}.  With the newly upgraded JVLA, it has been possible to begin performing dust polarization studies at centimeter wavelengths, using the highest-frequency bands available: K, Ka, and Q bands (18--50\,GHz).

\begin{figure} [hbt!]
\centering
\includegraphics[scale=0.11, clip, trim=0cm 0cm 0cm 0cm]{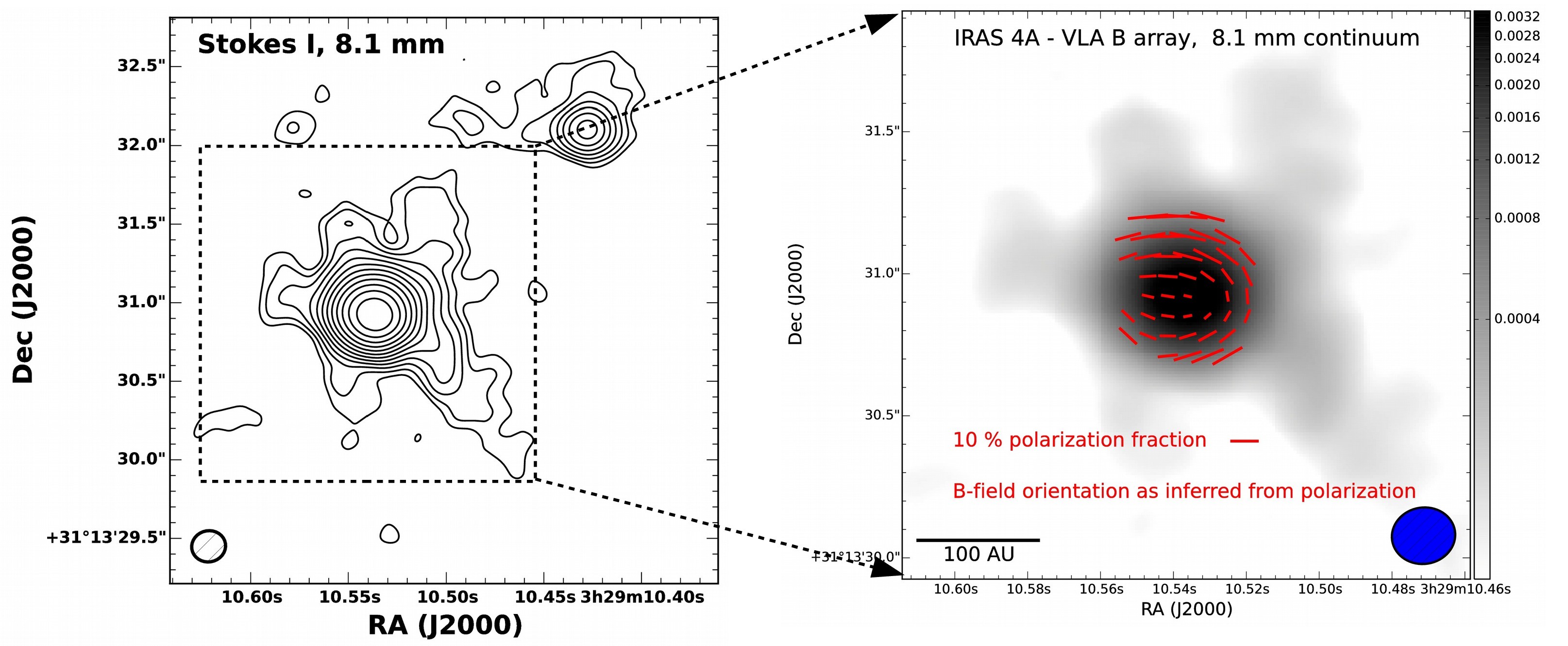}
\caption{\footnotesize
VLA B-config imaging of the deeply embedded protostar NGC 1333 IRAS 4A at 8.1\,mm (37\,GHz); Stokes $I$ emission is shown in the left panel and the magnetic field orientations inferred from the polarized emission (rotated by 90$\degree$) are shown in the right panel. The grayscale in the right panel is the Stokes $I$ emission. The rms noise is 23\,\ujybm{}; the contours start at 3\,$\sigma$ and increase in intervals of $\sqrt{2}$.  Credit: Erin Cox (Cox et al. 2015, submitted).
} 
\label{ch2_fig:IRAS4A_VLA_pol_cox} 
\end{figure}

\begin{figure} [hbt!]
\centering
\includegraphics[scale=0.33, clip, trim=0.4cm 1cm 2cm 0.5cm]{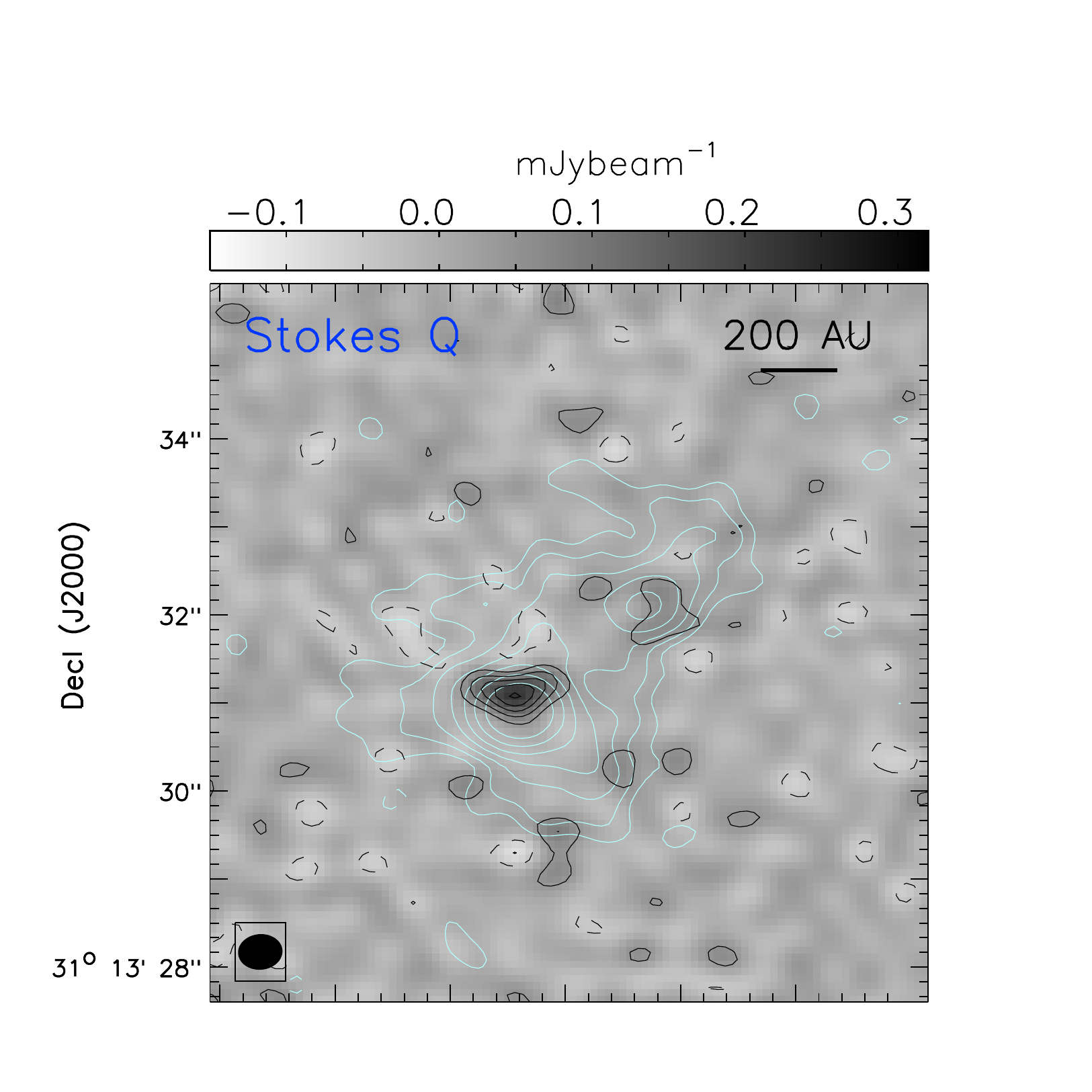}
\includegraphics[scale=0.33, clip, trim=3cm 1cm 2cm 0.5cm]{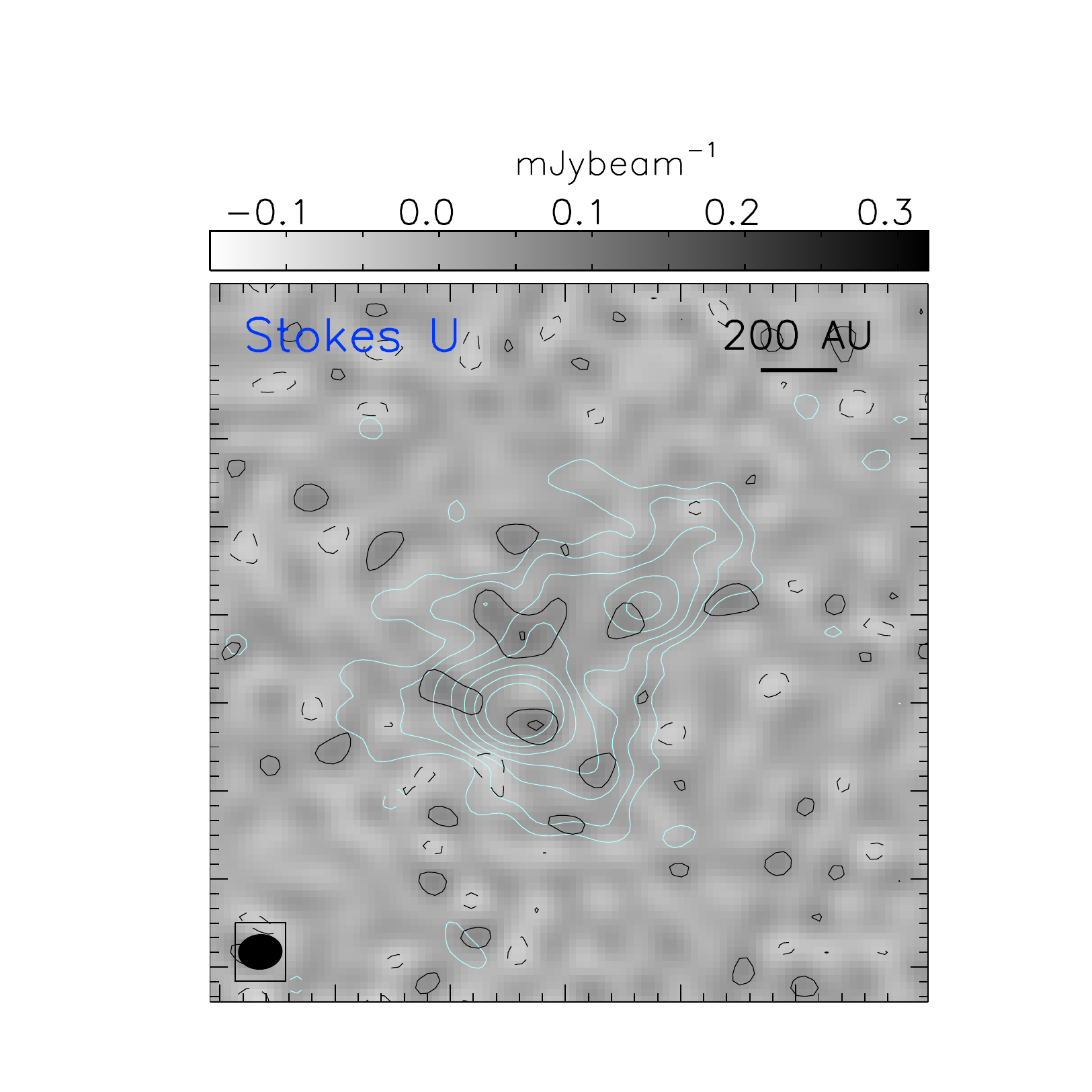}
\includegraphics[scale=0.33, clip, trim=3cm 1cm 2cm 0.5cm]{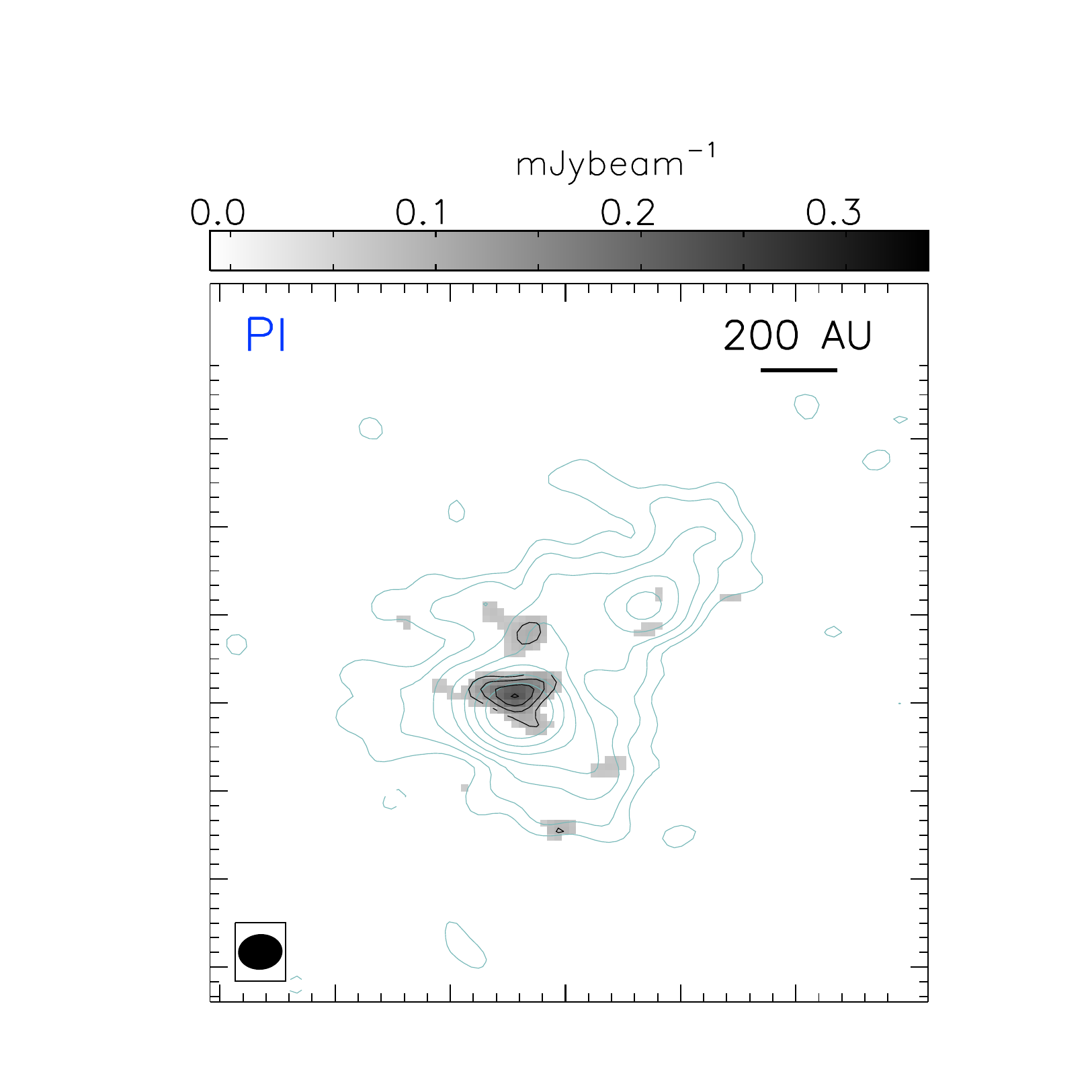}
\caption{\footnotesize
Observations of dust polarization toward the Class 0 young stellar object NGC 1333-IRAS 4A. % from VLA project 14B-053 (these maps are preliminary).  
From left to right are JVLA 6.9\,mm (43\,GHz) images of Stokes $Q$, Stokes $U$, and polarized intensity, produced with a synthesized beam of $\sim$\,0.44$\arcsec$.  The grayscale and the dark contours represent polarized intensity; cyan contours represent Stokes $I$ continuum emission.  Negative contours are dashed.  
The dark contours are --8, --6, --4, --2, 2, 4, 6, 8, 10 $\times$ $\sigma_P$, where $\sigma_P = 22$\,\ujybm{} is the rms noise in the map of polarized intensity.
The cyan contours are --6, --3, 3, 6, 12, 24, 48, 96, 192 $\times$ $\sigma_I$, where $\sigma_I = 19.5$\,\ujybm{} is the rms noise in the Stokes $I$ map.  
The polarization percentage at the peak of the VLA image is $\sim$2\%.
Credit: Baobab Liu (Liu et al. 2015, submitted).
} 
\label{ch2_fig:IRAS4A_VLA_pol_liu} 
\end{figure}

See Figures \ref{ch2_fig:IRAS4A_VLA_pol_cox} and \ref{ch2_fig:IRAS4A_VLA_pol_liu} for preliminary VLA detections of polarized dust emission at 37 and 43\,GHz towards NGC 1333-IRAS 4A, one of the first sources to be imaged in full polarization and high resolution at submillimeter wavelengths using the SMA \citep{Girart06}.  For the JVLA, NGC 1333-IRAS 4A is still one of the very few sources with bright enough polarized emission that it can be imaged within a few hours.  The improved sensitivity of the ngVLA will allow us to detect many more objects, enabling studies of long-wavelength polarization in a statistically significant sample of protostellar sources.  These studies will allow us to characterize the polarization in regions optically thick at ALMA wavelengths, and will also enable us to characterize the polarization properties of grains by observing the polarization spectrum (i.e., source polarization as a function of wavelength, from infrared to (sub)millimeter to centimeter wavelengths).

Considering that ALMA will eventually have Band 1 ($\sim$\,40\,GHz), and it already has a functioning polarization system at Band 3 ($\sim$\,100\,GHz), the lower-frequency K and Ka band (18--40\,GHz) observations would be unique to the ngVLA.  
Also, without considering the unknown polarization efficiency as a function of wavelength, 18--40\,GHz observations are easier than higher Q band (40--50\,GHz) observations at the JVLA site due to the weather constraints; this may not be an issue for the ngVLA depending on the weather at the site where the array is constructed.

\bigskip
\textit{Frequency requirement:}
The best frequency range to probe the polarized emission from large ($\sim$\,1\,cm) dust grains ranges from 20--50 GHz. 

\bigskip
\textit{Sensitivity requirement:}
An initial goal of the ngVLA would be to study the centimeter-wavelength polarization in the same sets of sources already observed at millimeter wavelengths.  In the high-resolution ($\sim$\,0.1$\arcsec$) 9\,mm VLA image of NGC 1333-IRAS 2A (Figure \ref{ch2_fig:multiplicity}) the peak flux density is approximately 1\,\mjybm{}.  To detect $\sim$1\% polarization in that source (and other similar sources in Perseus) at $> 5\sigma$ and at similar resolution would require an rms of $\sim$\,1\,\ujybm{}.

\bigskip
\textit{Angular resolution requirement:}
To match the resolution of previous millimeter observations will require resolutions of $\leq$\,3$\arcsec$; to observe polarization in the inner regions of disks where emission is optically thick at millimeter wavelengths will require a resolution of $\sim$\,0.1$\arcsec$.

\subsection{Structure of jets in deeply embedded protostars}
\label{ch3_jets}

\begin{figure} [hbt!]
\centering
\includegraphics[scale=0.6, clip, trim=0.5cm 2.5cm 0cm 0cm]{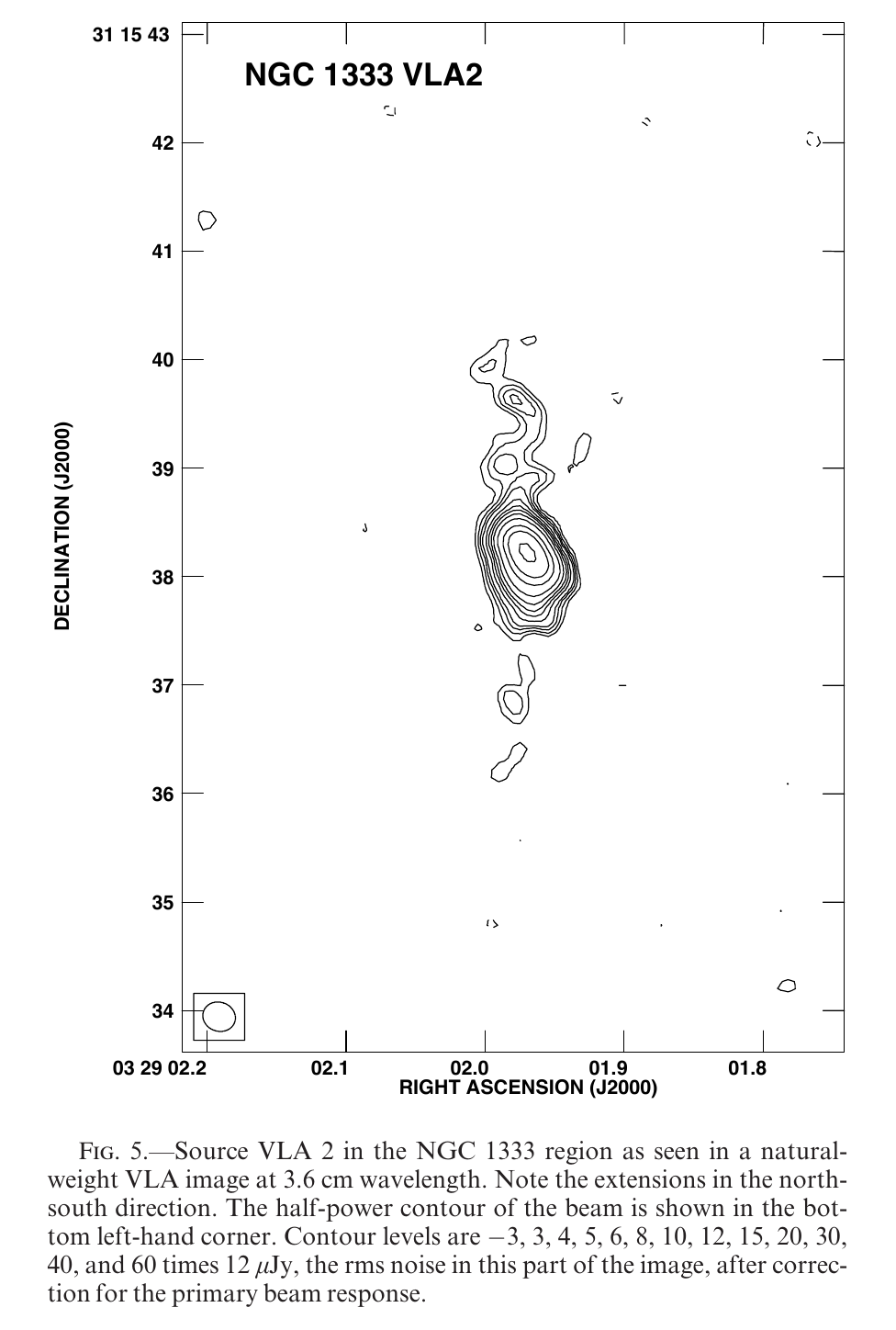}
\includegraphics[scale=0.7, clip, trim=0.5cm 2cm 0cm 0cm]{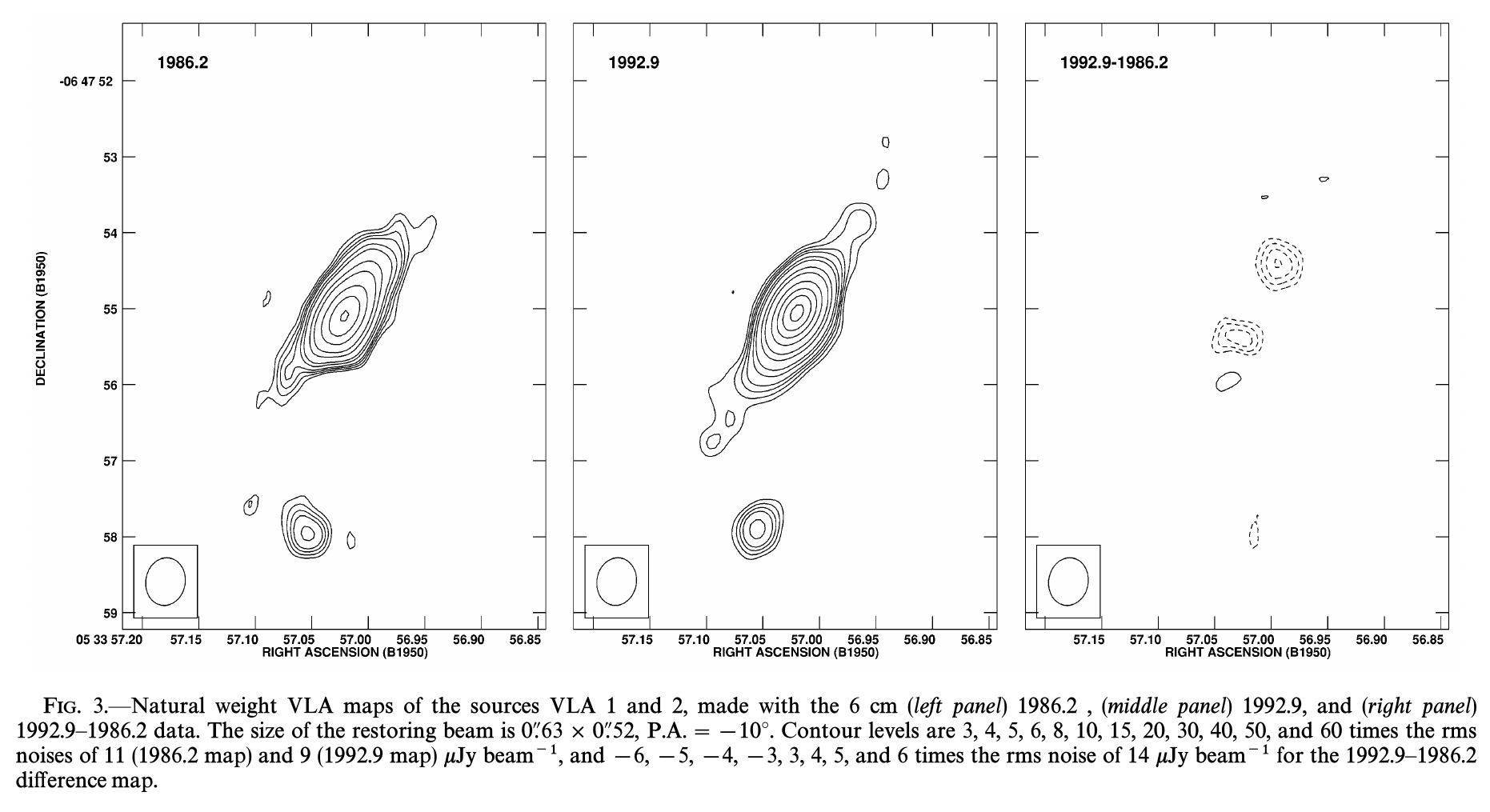}
\caption{\footnotesize
Examples of radio jets from protostellar sources.  \textit{Top:} NGC 1333-VLA 2 (commonly known as SVS 13C) from \citet[][Figure 5]{Reipurth2002}.  \textit{Bottom:} VLA 1, the driving source of the HH 1--2 flow, from \citet[][Figure 3]{Rodriguez2000}.
} 
\label{ch2_fig:radio_jets} 
\end{figure}

One of the key contributions of the JVLA to the field of low-mass star formation
was the discovery of compact free-free emission toward the central sources in low-mass protostars
\citep[e.g.,][]{Rodriguez1989, Anglada1998} and the strong correlation of free-free emission with
bolometric luminosity \citep{Shirley2007}. Some of these sources are found to be
extended, suggesting that powerful jets are being driven by low-mass protostars 
\citep[see Figure \ref{ch2_fig:radio_jets}:][]{Rodriguez2000, Reipurth2002}.

The JVLA has a maximum resolution at 12\,GHz of $\sim$\,0.2$\arcsec$ (28\,AU at 140\,pc);
at this resolution, the jet launching region is unresolved and much of the
free-free emission in the jets is optically thick \citep[e.g.,][]{Rodriguez2000, Shirley2007}. 
With a spatial resolution several times higher than that of the JVLA at
the relevant frequencies of $\sim$\,4--30 GHz, the
ngVLA will be able to zoom in on the jet-launching region and characterize the emission processes.
For example, observations at 12\,GHz will be able 
determine if the jet and/or outflow is being launched from an ionized disk wind and out to what
scales the jet is being launched (at 12\,GHz the ngVLA will have a resolution 
of $\sim$\,0.05$\arcsec$, which is 7\,AU resolution at 140\,pc).  By probing small spatial scales we can
also probe short timescales, which might allow us to see changes in brightness
(or even physical structure) of optically thin free-free or gyro-synchrotron
emission on orbital timescales.

The emission from protostellar jets is thought to be linked to the mass accretion process, 
where the outflow rate will increase as the mass accretion rate increases 
\citep[e.g.,][]{Pudritz2007}. Thus, variability in the protostellar jets could be used
as a diagnostic of variable accretion in low-mass protostars that are completely 
obscured at optical wavelengths. This avenue of exploring accretion variability will be 
extremely important given that most luminosity from protostars is emitted 
between 70\,$\mu$m and 160\,$\mu$m, and the space-based facilities needed 
to access these wavelengths have limited lifetimes. Finally, further observations 
are needed to constrain how strongly coupled accretion and outflow/jet emission
truly are \citep{GalvanMadrid2015}.

Additionally, one could use high-resolution observations of optically thin
lines or continuum to probe variability within the disk structure itself, which might
vary independently from the protostar. 

%%% BAOBAB'S ADDITIONAL VARIABILITY SUGGESTION:
%For the variability part of Section 3.2, you may refer to our JVLA studies
%on R Corona Australis (d~140 pc):
%http://adsabs.harvard.edu/abs/2014ApJ...780..155L
%
%For the sensitivity requirement, you may say a 10 times better sensitivity
%will (1) permit this type of studies to Orion molecular clouds, (2) can
%enable the time monitoring of the usually very radio faint Class II objects
%in the nearby (d~140 pc sources), (3) can enable the monitoring
%observations of the previously detected objects with the shorter than 1 min
%time cadence. and (4) enable the detection of the variability of linearly
%and circularly polarized gyrosynchrotron emission (typically polarization
%percentage is ~<20%).

\bigskip
\textit{Frequency requirement:} Jet emission will be observable and potentially
resolvable by the ngVLA at frequencies of $\sim$\,4--30 GHz.

\bigskip
\textit{Sensitivity requirement:} The jet emission from a 1\,L$_{\odot}$ 
protostar at a distance of 140\,pc typically has a flux density of 300\,$\mu$Jy at 12 GHz \citep{Shirley2007}.
If this emission is coming from an ionized disk wind from an $R=50$\,AU disk, the surface brightness
will be $\sim$\,6\,\ujybm{}. To detect this with a SNR of 10, a sensitivity of 0.6\,\ujybm{} is necessary.

\bigskip
\textit{Angular resolution requirement:} In order to begin resolving structure within free-free
jets, spatial resolutions several times higher than the JVLA resolution will
be required at frequencies of $\sim$\,4--30\,GHz: i.e., $\sim$\,0.05$\arcsec$ resolution at 12\,GHz. Higher
frequencies will also enable the jet emission to be isolated with higher resolution.

\clearpage
\section{Formation of planets}
\label{ch4}

\subsection{Mapping planet formation regions in nearby disks}
\label{ch4_KSP}

Planets are predicted to form in the innermost, densest regions of dusty and gaseous disks orbiting young 
pre-main sequence stars. Due to the high density and optical depth of these regions, observations at 
millimeter and centimeter  wavelengths are required to 
study the physical processes that control the formation and early evolution of planetary systems.

\begin{figure} [t!]
\centering
\includegraphics[scale=0.52]{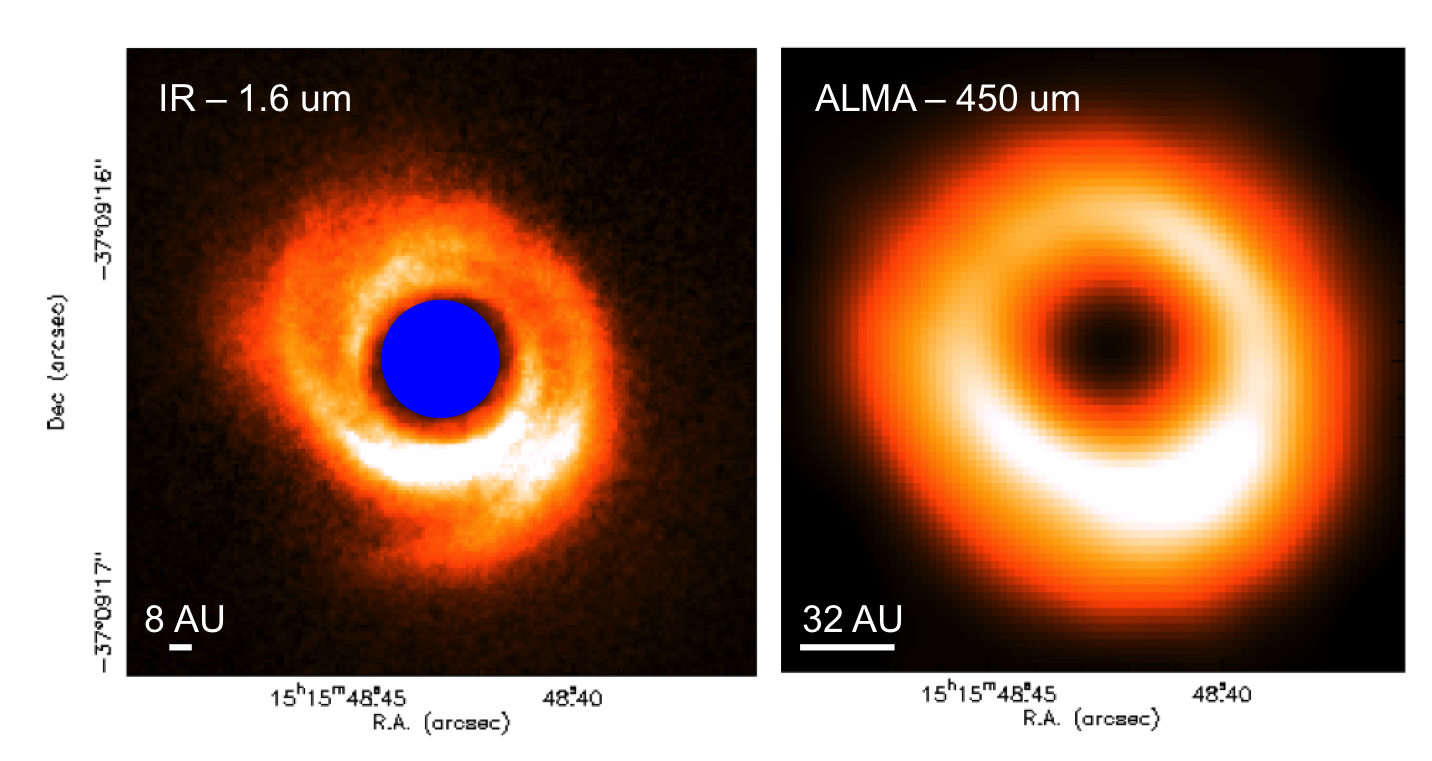}
\caption{\footnotesize  Map of the dust emission observed toward the SAO~206462 young stellar system. \textit{Left:} image of the near-infrared
scattered light emission observed with HiCIAO at Subaru \citep{Muto2012}. \textit{Right:} image of the 450\,$\mu$m 
dust continuum emission observed with ALMA \citep{Perez2014}. The angular resolution of the observations is indicated on 
the lower left of each panel.}
\label{fig:sao206462} 
\end{figure}

High-resolution observations of the (sub)millimeter-wave emission from young 
circumstellar disks have revealed large holes and asymmetries in the circumstellar 
dust and gas distributions, which are sometimes accompanied by spiral structures observed 
at near-infrared wavelengths \cite[see Figure~\ref{fig:sao206462} and the review by][]{2014prpl.conf..497E}. 

These structures can be interpreted as the result of the gravitational interaction between as-of-yet 
unseen giant planets and the circumstellar material, and provide 
information on the mass and orbital radius of newborn planets. The most outstanding example is the discovery of 
annular gaps in the circumstellar dust distribution around the young ($< 10^6$ yr) low mass star HL Tau (see Figure \ref{ch1_fig:HLTau}), 
which suggests the presence of giant planets orbiting at tens of AU from the central star. 

A major limitation of observations at wavelengths shorter that 1--3\,mm is that they are incapable of penetrating 
dust column densities higher than about 1--3 g cm$^{-2}$ (assuming a dust opacity as in Table \ref{table:optical_depth}), therefore 
making impossible to study the formation of planets in the innermost regions of disks.  
For example, the innermost 10 AU of a Minimum Mass Solar Nebula disk would be opaque 
at the wavelengths covered by ALMA.  This is presumably where the planets 
in our Solar systems formed, and, in general, where the majority of the planets are 
expected to form. 

%, where the dust column density is predicted to exceed this value 
%(for example, in a MMSN disk model, $\Sigma_d > 1$ g cm$^-2$ for $r < 6$~AU).

\begin{figure} [t!]
\centering
\includegraphics[scale=0.53]{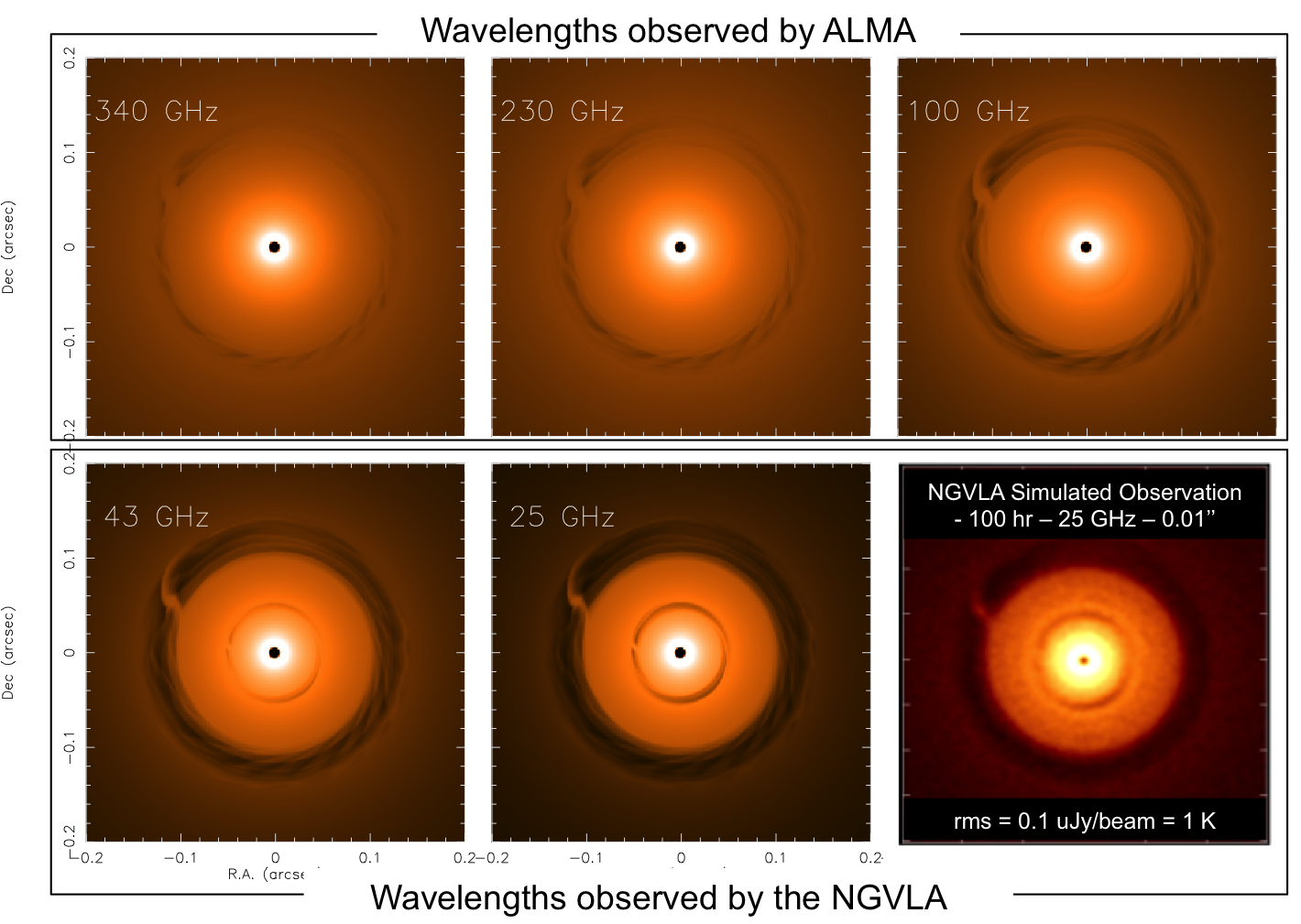}
\caption{\footnotesize  Multi-frequency maps of the synthetic continuum emission from a planet forming disk located at the 
distance of 130 pc from the Earth.  The disk hosts a Jupiter and a Saturn mass planet at 12 AU and 6 AU from the central 
star, respectively. Because of the high optical depth within a radius of 10 AU, 
the emission at frequencies >\,50 GHz reveals only the outermost planet.  
The emission at frequencies <\,50 GHz reveals both 
planets. The bottom right panel shows the simulated ngVLA observation of the 25\,GHz disk emission. The angular 
resolution of the ngVLA observations is 10 mas and the noise level is 1\,K. }
\label{fig:psys5} 
\end{figure}

By mapping the disk emission at centimeter wavelengths, the ngVLA offers a unique opportunity 
to investigate the distribution of dust and gas in the innermost disk regions not accessible to ALMA. 
An example of the imaging capabilities provided by the ngVLA is shown in Figure~\ref{fig:psys5}.
ngVLA observations of the dust continuum emission between 1--3\,cm will penetrate dust layers 
as thick as 12 g cm$^{-2}$, which, in a MMSN disk, corresponds to an orbital radius of about  1 AU. 

Last but not least, by accessing the innermost disk regions, multi-epoch ngVLA observations  will  allow us 
to follow the temporal evolution of planet forming disks on time scales comparable with the 
orbital period of the circumstellar material. 

%The requirement to map the dust continuum emission at sub-AU resolution implies that it will be necessary to achieve 
%a sensitivity of about 0.5 K (this corresponds to a S/N > 10 assuming a disk midplane temperature > 50 K and an optical depth of 0.1). 
%This is achieved by the preliminary ngVLA design, which delivers a sensitivity of 0.32 K with 182 km baselines (i.e., 10 mas at 1 cm) 
%and 10 hours of integration on source.\\

\textit{Frequency requirement:} Frequencies between 10 GHz and 50 GHz are ideal to map the innermost regions of protoplanetary 
disks because they provide the best combination of optical depth, intensity of the dust emission, and angular resolution 
that can be achieved with the ngVLA.   \\

\textit{Sensitivity requirement:} A sensitivity of 0.5--1\,K is required to map the dust emission with a peak signal-to-noise ratio  
larger than 10 (this assumes a disk midplane temperature of 50\,K and an optical depth of 0.1). \\

\textit{Angular resolution requirement: } An angular resolution of 5--10 mas is required to detect the perturbations of forming planets that lie within 10 AU 
of nearby ($d = 130$\,pc) young stars. \\

\textit{Spatial filtering requirement:} a maximum angular scale of about 5$\arcsec$ needs to be mapped in order to recover the total disk flux. \\

\subsection{Probing planet formation through the distribution of dust and pebbles}
\label{ch4_sec:dust}

The spatial distribution of circumstellar dust is shaped by the aerodynamic interactions between dust and gas, which strongly depend on the dust 
grain size. In particular, grains smaller than a few microns are coupled to the gas, while millimeter dust grains and larger 
are concentrated toward gas pressure maxima. This process can lead to local enhancements in the dust-to-gas ratio that have 
strong implications for the formation of planetesimals \citep[see, e.g., the PPVI review by][]{Johansen2014}.

Before planet formation occurs, large dust grains settle toward the disk midplane and migrate toward the innermost disk regions, where they interact to form 
planetesimals. However, as soon as the first planets form, they gravitationally interact with the circumstellar gas, producing 
local enhancements in the gas pressure that can catalyze the formation of planetesimals at much larger orbital radii. This 
process is currently observed in transitional disks and might have played an important role in forming trans-Neptunian 
objects and comets in our Solar System.

\begin{figure} [t]
\centering
\includegraphics[bb= 120 80 870 900, clip=True, scale=0.15]{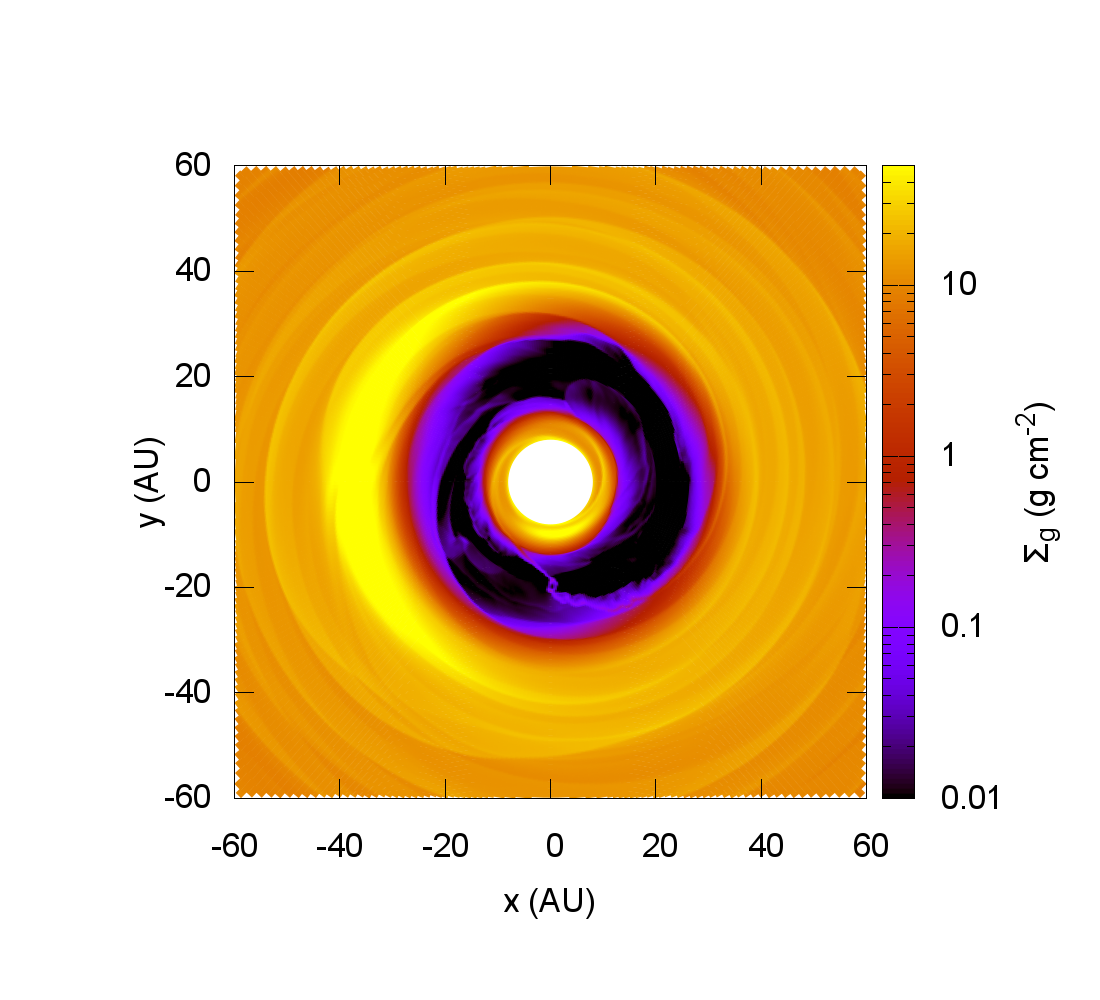}
\includegraphics[bb= 220 80 870 900, clip=True,scale=0.15]{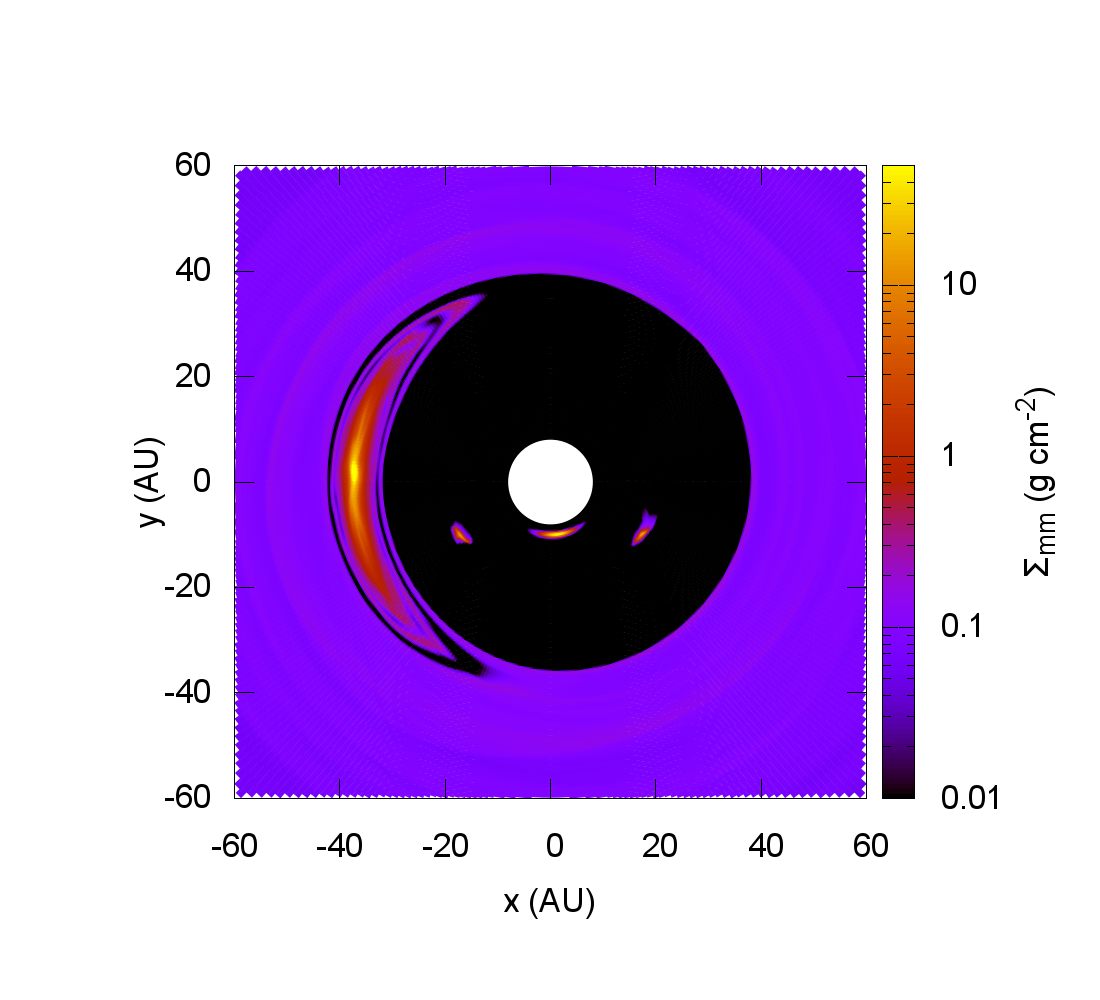}
\includegraphics[bb= 220 80 1100 900, clip=True,scale=0.15]{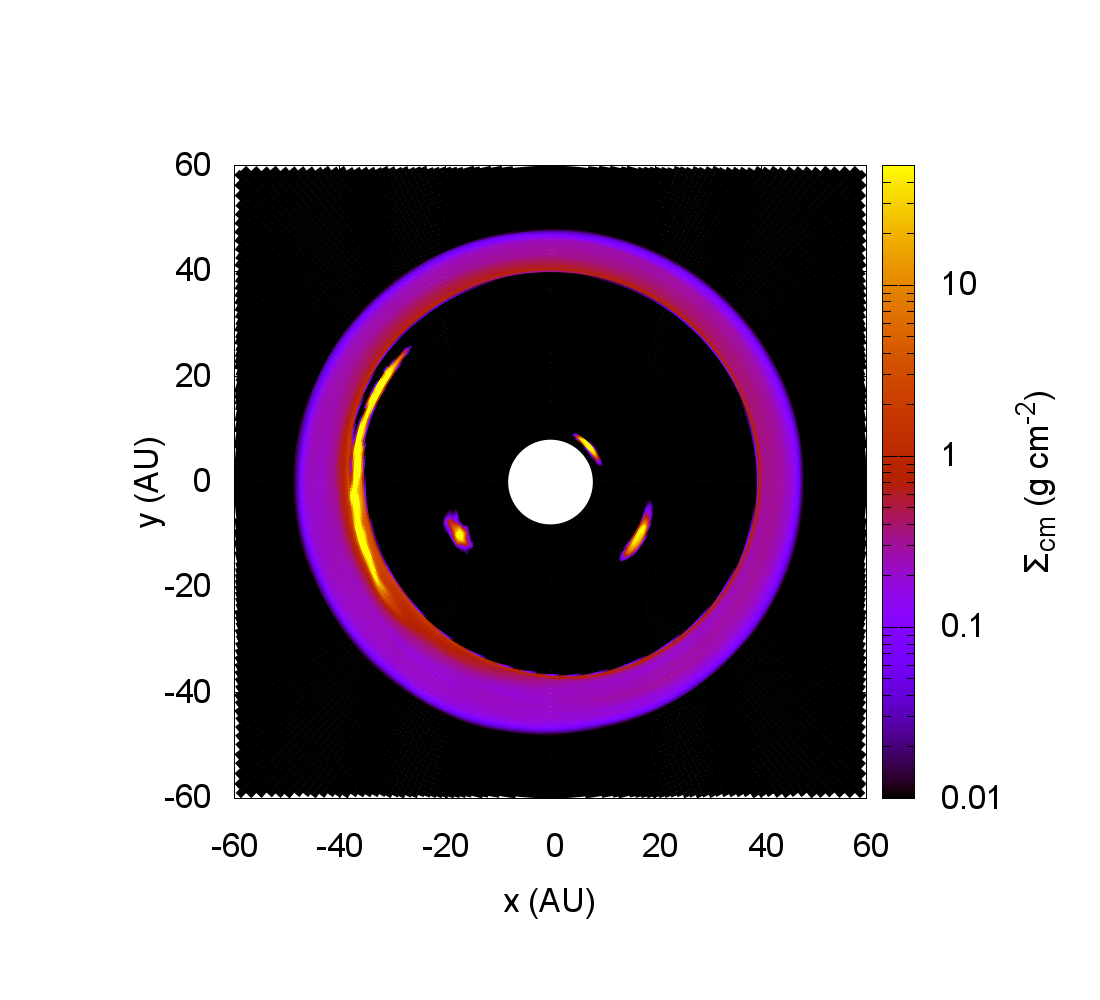}
\caption{\footnotesize Snapshot of the gas and dust surface density of a disk perturbed by a 10 M$_J$ planet orbiting at 20 AU from the central star, after 600 orbits of the planet. The planet is located at (0, --20). 
\textit{Left:} Gas surface density. A circular gap and a vortex develop at the orbital radius of the planet and at the outer edge of the gap, respectively. \textit{Center:} Surface density of 1 mm grains. Millimeter grains are concentrated toward the center of the vortex (where the dust-to-gas ratio reaches values close to 1), and in the Lagrangian points on the planet orbit. \textit{Right:} Surface density of 1 cm grains. Centimeter grains are even more concentrated toward the center of the vortex \citep[modified from][]{Fu2014}. } 
\label{fig:disks} 
\end{figure}

The ngVLA will allow us to map the distribution of pebbles across the entire planet formation process, and
the comparison between ALMA and ngVLA observations will allow us to investigate the evolution of solids and 
the formation of planetesimals (see Figure~\ref{fig:disks}).\\

\textit{Frequency requirement:} Frequencies between 10--50\,GHz are required to map the distribution of centimeter-size pebbles. \\

\textit{Sensitivity requirement:} A sensitivity of 0.5--1\,K is required to map the dust emission with a peak signal-to-noise ratio  
larger than 10 (this assumes a disk midplane temperature of 50 K and an optical depth of 0.1). \\

\textit{Angular resolution requirement:} An angular resolution higher than 0.2$\arcsec$ is required to spatially resolve the distribution of pebbles in 
circumstallar disks at a distance of 130\,pc. An angular resolution higher than 0.1$\arcsec$ is required to spatially resolve the 
sub-structures predicted by dust evolution models (see Figure~\ref{fig:disks}). An angular resolution higher than 0.05$\arcsec$ is 
required to spatially resolve the vertical distribution of dust.\\

\textit{Spatial filtering requirement:} a maximum angular scale of about 5$\arcsec$ needs to be mapped in order to recover the total disk flux. \\

\clearpage
\section{Chemical evolution of the star and planet forming environment}
\label{ch4.5}

\subsection{Complex organic molecules in Hot Molecular Cores}

Hot molecular cores (HMCs) are the best reservoirs in the Galaxy of complex organic molecules
(COMs).  These COMs include prebiotic molecules, which are the building blocks of life. 
Prebiotic molecules such as glycolaldehyde or ethylene glycol were
first detected in the interstellar medium towards Sgr B2(N) in the Galactic Center 
\citep[e.g.,][]{Hollis02,Hollis00}, and subsequently in HMCs such as G31.41+0.31
\citep{Beltran09}, Orion KL \citep{Brouillet15}, and in the hot corino around the 
Solar-type protostar IRAS 16293-2422 \citep{Jorgensen2012}. COMs are important
not only because of the link between prebiotic molecules and the origin of life, but also because COMs
can be used to study higher density
regions than those traced by typical hot-core tracers such as methyl cyanide
\citep[e.g.,][]{Beltran09}. This allows us to glimpse deep inside
molecular cores and study the material closer to the central (proto)star, where
a potential massive circumstellar disk might be embedded.

\sloppy{COMs in high- and low-mass star-forming regions can be easily observed at (sub)millimeter}
wavelengths but are severely affected by line blending. Moving to the centimeter-wavelength
window, the number of spectral lines in a selected spectral portion is
significantly lower with respect to the millimeter window, and hence the lines are less
blended. This allows us to more easily detect the presence of COMs, and
to study the kinematics of high- and low-mass star-forming regions---in particular of
possible accretion disks---in a very pristine way \citep[e.g.,][]{Goddi15, Codella2014}.

\bigskip
\textit{Frequency requirement:} COMs have several transitions at cm wavelengths,
but to achieve the higher angular resolution needed to trace the inner regions
of HMCs a good compromise would be to observe between 18--50\,GHz.

\bigskip
\textit{Sensitivity requirement:} We use glycine as a benchmark due to its high prebiotic interest. Based on the predicted hot core glycine abundances from \citet{Garrod2013} we can expect line brightnesses of $\sim$\,0.15\,K for the strongest lines in the nearest HMCs, assuming a hot core temperature of 200\,K and a spatial resolution of 0.2$\arcsec$.

\bigskip
\textit{Angular resolution requirement:} At the typical distances of high-mass star-forming
regions ($\sim$\,5 kpc), one needs angular resolutions of $<0.1\arcsec$ to trace HMCs,
which have typical sizes of a few 1000\,AU (0.02--0.04\,pc). However,  to
trace possible embedded disks with spatial scales of a few 100\,AU (typical of
disks around low- and intermediate-mass protostars) one needs angular
resolutions of 20--50 mas, easily achieved by the ngVLA between 27--50\,GHz.

\bigskip
\textit{Spectral resolution requirement:} Typical line widths of COMs in HMCs are of a few km/s.

\subsection{Gas and chemistry in planet forming regions}
\label{ch4_sec:chemistry}

Investigating the chemistry of planet forming regions in young circumstellar disks is key for understanding 
the initial condition that lead to the development of life.  The primordial amount of carbon, nitrogen, oxygen, and water
present on planets depends on the molecular composition of the parent disk and on the planet formation 
history. Furthermore, how far the chemistry of complex organic molecules and prebiotic molecules extends into 
the circumstellar and interstellar media needs to be understood in order to determine the complexity 
of the material that is delivered to planetary surfaces.

The low frequencies of the ngVLA (compared with ALMA) give access to:

\textit{Ammonia (NH$_3$),} one of the most (perhaps the most) important N-bearing volatile. NH$_3$ is a major carrier of nitrogen in interstellar and perhaps circumstellar environments. It is also the most commonly used gas thermometer. 
NH$_3$ is frequently observed in pre- and protostellar regions, but lack of sensitivity has prevented its detection in protoplanetary disks. Obtaining 
accurate gas temperatures in disks is a key goal of disk explorations, since temperature regulates both the chemistry and the physics, including planet migration 
patterns. The importance of nitrogen for prebiotic chemistry also makes it disturbing that we currently do not have observational access to any of 
its main carriers during planet formation. Using the chemical model described in Du et al. (2015), we calculate that detecting ammonia in disks is within the 
reach of ngVLA observations. In particular, we estimate that brightest ammonia radio wavelength transitions around 23.7 GHz from TW Hya disk 
can be detected with a signal-to-noise larger than 5 in 1 hour on source.  Detecting the same transitions in disks at a distance of 130 pc would therefore 
require about 10 hrs of integration on source. 

\textit{Volatile molecules in very dense regions,} e.g., the innermost parts of protoplanetary disks, that are otherwise veiled by optically thick dust at shorter wavelengths. 
With the advent of high-spatial resolution and high-sensitivity molecular line millimeter observations with ALMA, it is becoming clear that a major obstacle to exploring the disk midplane of the innermost 10s of AU of disks is dust optical depth. This can only be resolved by high-spatial-resolution observations at longer wavelengths where the dust opacities are smaller. Since this region is the planet forming region in most disks, characterizing its volatiles is an important objective. Accessible ngVLA lines include NH$_3$ (see above), H$_2$CO(1--1, 2--2, 3--3) CH$_3$OH(1--0), and perhaps OH, HCO$_2^+$, HDO and H$_2$O if the ngVLA is designed with sufficient sensitivity. 

\textit{Low-lying excitation modes of moderately complex organic molecules,} enabling the observations of molecules such as CH$_3$CN in the cold regions where they are proposed to form through ice chemistry. Most observations of COMs have focused on hot regions where thermal desorption of icy grain mantles bring all volatiles into the gas phase (e.g., \citealt{Oberg2015} have detected warm CH$_3$CN in a protoplanetary disk). These molecules form at low temperatures, however, and tracing the chemical evolution requires observations of low-lying excitation states of the small abundances of organic molecules that desorb non-thermally in their formation zones. The centimeter and 7\,mm spectral regions give access to these transitions for all common complex organic molecules, including CH$_3$CN.

\textit{Lines of very complex organic molecules such as glycine,} because of a lower line density at centimeter wavelengths compared to millimeter wavelengths. Very complex organics are difficult to detect for two reasons: (1) they are less abundant and therefore require high sensitivity, and (2) at these low signal strengths the millimeter spectra are crowded with emission from complex molecules and their isotopologues. At longer wavelengths the spectra are less crowded, enabling their detections given enough sensitivity. The recent detection of water vapor in the prestellar core L1544 shows that even in the cold interior of dense cores, cosmic rays yielding secondary high energy radiation can desorb a measurable quantity of molecules from the ices \citep{Caselli2012}. \cite{Jimenez2014} have shown that not only water, but also complex prebiotic molecules like glycine may be detectable if desorbed together with the water molecules. A similar argument could be made for the cold midplanes of protoplanetary disks.

\bigskip
\textit{Frequency requirement:} Frequencies from 10--40\,GHz are required to map the emission from the molecules described above.
A frequency resolution between 0.1--1\,km is required to resolve the line profile and study the disk kinematics.

\bigskip
\textit{Sensitivity requirement:} Given the large uncertainties in molecular abundances, it is extremely difficult to predict the emission from complex 
molecules. A sensitivity of about 30\,\mjybm{} in a 1\,\kms{} channel is required to detect ammonia in disks. 

\bigskip
\textit{Angular resolution requirement:} An angular resolution between 0.2--5$\arcsec$ is required to detect, and possibly spatially resolve, 
molecular line emission from circumstellar disks. 

\bigskip
\textit{Spatial filtering requirement:} A maximum angular scale of about 5$\arcsec$ needs to be mapped in order to recover the total disk flux.

% Debris disks
\clearpage
\section{Structure of Debris Disks}

The end stage of circumstellar disk evolution is the debris disk, where dust 
is produced by the collisional erosion of planetesimals.
% , analogous to comets and asteroids.  
Imaging the structure of debris disks is important to 
indirectly infer the presence (and perhaps the orbital evolution) of planets
on wide orbits that are not accessible via other observational techniques. 
While the kilometer-sized remnants of the planet formation process cannot be
detected directly, emission from their dusty debris is linked to them through 
the collision process and offers a window into their physical and dynamical
properties.  Observations of debris disks at millimeter wavelengths are 
important because the large grains that dominate the thermal emission 
directly trace the planetesimals, unlike the smaller grains seen in the optical 
and infrared that are rapidly blown out by stellar radiation and winds. 

Because of the low masses of dust in debris disks compared with the younger
primordial disks, and because dust opacities drop steeply with wavelength, 
only a few detections of debris disks in dust continuum emission at 
$\lambda > 3$\,mm have been reported \citep[see, e.g.,][]{Ricci2012}.
In general, ALMA observations at shorter wavelengths can probe the structure
of small solids (and therefore planetesimals) at higher signal-to-noise 
ratios than the ngVLA. However, the improved sensitivity of the ngVLA will
enable the detection of many more debris disks at longer wavelengths, where the spectral 
index of the dust continuum emission can be related simply to the underlying 
size distribution of centimeter-sized solids, thereby probing the collisional 
models.  

The reference model is the steady-state catastrophic collisional 
cascade, where the relative velocities and tensile strengths of the colliding 
bodies are assumed to be independent of size, and velocities are high enough 
to create fragments that are all less than half the size of the original
parents. This self-similar shattering recipe leads to a power-law size 
distribution, $n(a) \propto a^{-q}$ with exponent $q=-3.5$. However, the 
process is certainly more complex, and recent analytic and numerical studies
incorporate more realistic collision dynamics and physics, and generally 
predict steeper size distributions \citep[e.g.,][]{Pan2012}. It is also 
possible that the particles are more like rubble piles, or that velocities are 
small and collisions are gentle, which would lead to a more shallow size 
distribution (e.g., Saturn's rings).

Observations of debris disks with the ngVLA, together with ALMA, will probe,
via the emission spectral index, the size of the solids involved in the collisional cascade. 
Observing with the ngVLA is advantageous because the effect of 
dust temperature (which also affects the spectral index) is mitigated 
at longer wavelengths.  In addition, ngVLA observations of debris disks
will enable us to investigate whether grains with different sizes have different 
spatial distributions, and, ultimately, to look for clumps of centimeter size grains retained in 
resonance with unseen planets (see, e.g., \citealt{Wyatt2006}).  

Finally, longer wavelengths also provide a long
lever arm to minimize the effects of absolute calibration
uncertainties. Since it seems likely that not all debris disks are the same,
and that solid properties may vary with dynamical history and radiation 
environment, large samples of debris disks will need to be observed to reach
general conclusions about collisional processes in mature planetary systems. \\

\textit{Frequency Requirement:} Frequencies from 10--50\,GHz provide good sensitivity
to thermal dust emission from large grains. An extension to 100\,GHz would be 
useful to better define the spectral shape. \\

\textit{Sensitivity Requirement:} To reach a sample of 10s to 100s of debris disks
requires an rms sensitivity of roughly 100 \njybm{}.\\

\textit{Resolution Requirement:} A beam size of a few arcseconds resolution is adequate
to obtain flux densities of debris disks; only the most nearby targets
will be resolved.

\clearpage
\section{Solar System}
\label{ch6}

The frequency range that may be covered by the ngVLA is appropriate for measuring
radiation from all categories of Solar System bodies: terrestrial and gaseous
planets, rocky and icy small bodies (moons, asteroids, Trans-Neptunian Objects
and other small-body families), and comets.  The long-wavelength tail of the
thermal emission from these bodies can be measured, allowing one to detect and
in many cases map the brightness temperature fields from surfaces or
atmospheres. The continuum emission (or radar reflection) spectrum originates
from depths in subsurfaces or atmospheres that cannot be probed by other
techniques, as well as from relatively large grains in planetary environments
(rings). The combination of long-wavelength measurements with observations at
shorter wavelengths can hence provide important and unique insight into
vertical atmospheric profiles, grain size distribution and, for surfaces,
subsurface structure. In addition, spectroscopic observations of molecular lines
(such as NH$_3$, CO or OH) can provide information on gas abundances and
temperatures in atmospheres and cometary comae. If the covered frequencies of
the ngVLA extend low enough (below 10 GHz), then non-thermal synchrotron
emission from electrons trapped in the magnetospheres of the giant planets can
be mapped, yielding important information on the distribution of the electrons
and the topology of the magnetic fields.

Strong constraints can be placed on a large range of properties
of Solar System bodies by such measurements, which are highly complementary to ground- and space-based observations in the optical, infrared and (sub)millimeter regimes.  The combination
of the exquisite sensitivity, excellent imaging performance, and especially high spatial resolution that could be offered by the ngVLA will expand the reach
and precision of the type of studies that are now only partially addressed by
the JVLA.

We present here a non-exhaustive list of scientific cases that can be addressed
through ngVLA observations. An extensive review of the science cases that can
be addressed by SKA, partially overlapping the ngVLA cases at the lowest frequencies, can be found in
\citet{butler2004}.

\subsection{Deep atmospheric mapping of giant planets}
\label{ch6_KSP}

Understanding the coupling of gas abundances, temperature, and dynamics of the deep
atmospheres of the giant planets is vital to our understanding of these planets
as a whole, and to our understanding of extrasolar giant planets.  Optical and
infrared observations probe only to depths less than 1 bar, so it is not known whether
the cloud features (including storms) seen at those wavelengths extend into the
deep atmospheres of the planets.  Observations at ngVLA wavelengths will probe
to much greater depths than the optical or the infrared---to tens of bars on Saturn, Uranus, and
Neptune and a few bars on Jupiter (deeper if frequencies below 10\,GHz are
available)---and will yield information on atmospheric properties' horizontal and vertical distributions, and their variations.
Previous observations at these wavelengths from the JVLA have provided important
information \citep{sault2004,depater2014}, but the sensitivity and resolution of the ngVLA will be so
superior that such observations will revolutionize our understanding of the deep
atmospheres of these planets. For example, on Neptune and Uranus (2.2$\arcsec$ and 3.6$\arcsec$ diameter), only the largest-scale features in the troposphere can be detected with the JVLA. The most extended configurations of the ngVLA would enable mapping of features with a characteristic size down to 450~km on Neptune.

\begin{figure} [hbt!]
\centering
\includegraphics[scale=0.6, clip, trim=0cm 0cm 0cm 0cm]{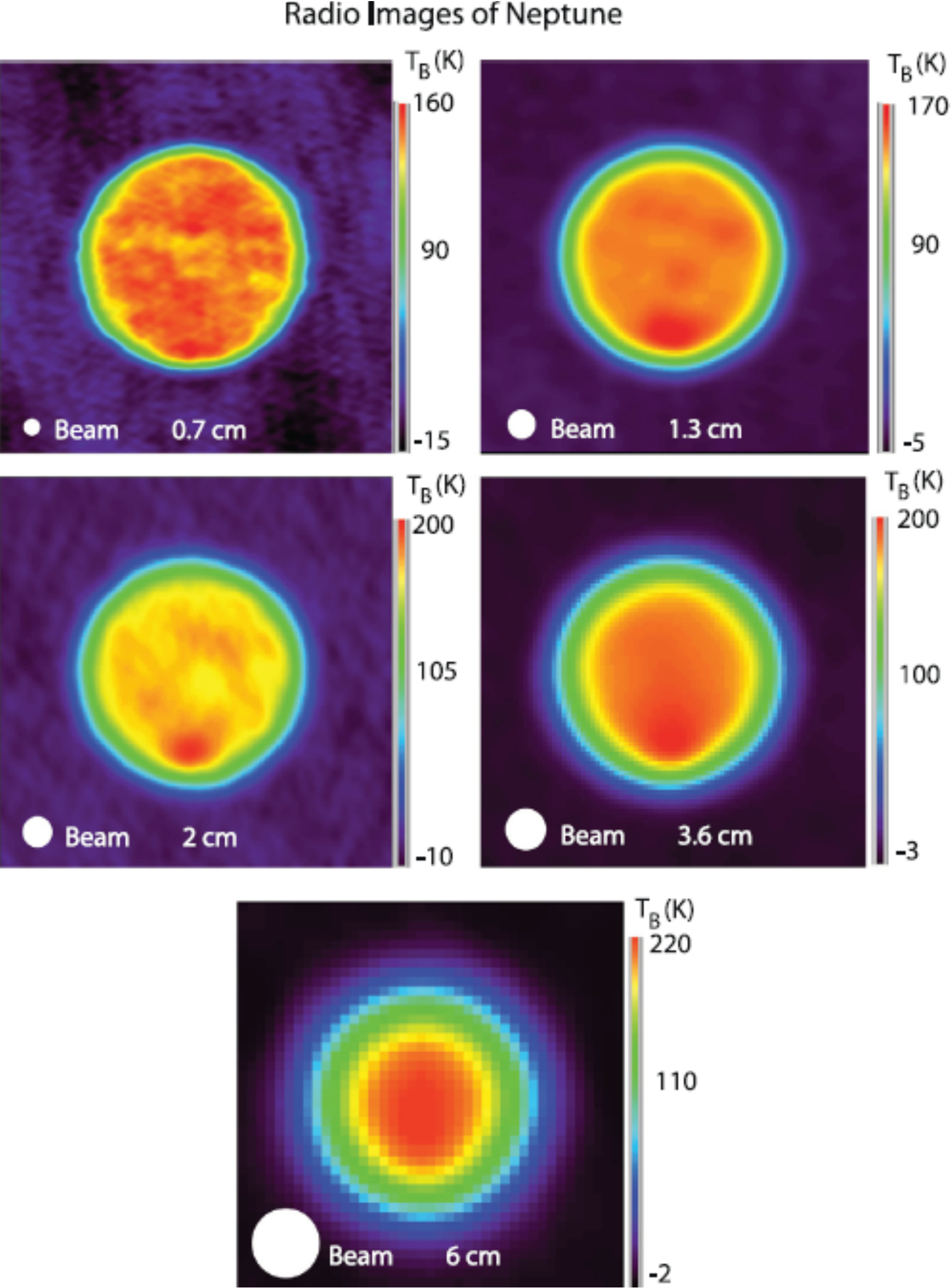}

\caption{\footnotesize JVLA radio images of Neptune at five wavelengths, as indicated. The intensity scale in units of brightness temperature is indicated on the right for each image, and the beam size (FWHM) is indicated in the lower left corner. Neptune has been rotated in these images so that the north pole is pointing up.  Reproduced from \citet{depater2014} (Figure 6).
} 
\label{ch6_fig:depater} 
\end{figure}

At ngVLA wavelengths, the bulk of the emission from the atmospheres of the giant
planets results from collisionally induced absorption of the individual
atmospheric molecules (H$_2$, NH$_3$, PH$_3$, H$_2$S, H$_2$O, etc.).  These
molecules do not form observable spectral lines, because the pressures are so
great that they are pressure-broadened into a pseudo-continuum.  If the CO(1--0)
transition is available to the ngVLA, the emission comes from high enough in the atmosphere that a more
typical spectral line is seen, at least on Neptune \citep{marten1993}.  By observing over a
large range of radio frequencies, one can probe different depths in the
atmospheres, and can hence determine the 3D spatial distribution of these
absorbers in a planet's deep atmosphere, as well as the absorbers' coupling to the vertical
temperature profile.  For example in \citet{depater2014}, VLA observations
were used to determine the humidity of the deep troposphere near Neptune's South
pole (see Figure \ref{ch6_fig:depater}).

A difficulty with the giant planets is their rapid rotation (of order 10 hours
for all of them), which can prevent the identification of longitudinal features in the atmosphere.  We cannot simply observe for longer periods to gain sensitivity and to allow for Earth rotation synthesis to fill in uv-coverage
for better imaging of these bright and complex sources, because they are
changing significantly with time.  A technique to overcome this difficulty, and
construct longitude-resolved images, has been developed by \citet{sault2004}.
When applied to VLA observations of Jupiter, it allowed for the first time a
direct comparison of features seen in the optical/infrared to those in the radio (at 1.5\,cm
wavelength), showing a very good correlation between radio and infrared
features.  This technique has been applied using more sensitive JVLA data
\citep{depater2014j}, which shows even finer detail than the previous
observations.  In principle the technique can also be used on the other giant
planets, but that has not been demonstrated yet.  For this technique to work
best, good instantaneous uv-coverage is required over the full range of spatial
scales of the planet---from the whole disk (short spacings) down to the
synthesized beam scale (long spacings).  Only with an interferometer like the
ngVLA is this possible.

\bigskip
\textit{Frequency requirement:} The frequency coverage determines the range of depths probed in the atmosphere. Since centimeter-wavelength mapping is the only technique that allows us to directly probe the deep atmosphere, the widest coverage is desirable.

\bigskip
\textit{Angular resolution requirement:} Any improvement in spatial resolution with respect to JVLA would directly translate into the ability to observe smaller atmospheric features. 

\bigskip
\textit{Field of view requirement:} A large instantaneous field of view ($\sim$\,1 arcmin) is preferable for imaging giant planets. 

\bigskip
\textit{Imaging requirements:} Observations on short and long baselines are essential to capture the large range of spatial scales at play, and must be obtained simultaneously due to the relatively short timescale of dynamic atmospheric features.

\bigskip
\textit{Other requirements:} Studies of time-variable phenomena may require flexible scheduling capabilities.\\

\subsection{Surface and subsurface thermal emission}

Since surfaces are not opaque at radio wavelengths, the thermal emission from
bodies where a surface can be sensed represents the contribution from not only
the surface, but from the subsurface as well. The rule of thumb for rocky
surfaces is that the main contribution comes from depths within about $\sim$10
wavelengths of the surface. At ngVLA wavelengths, this would correspond to
sounding depths in the 3--60 cm range. Such a wide range is of particular
interest since it samples both within and beyond the typical diurnal skin depth
(section of surface sensitive to diurnal variations). The possibility to measure
the vertical temperature profile across the skin depth provides strong
constraints on surface properties such as thermal inertia, and also reveals the
nature of the surface (porosity, roughness). Obtaining continuum spectra of
spatially unresolved small bodies (asteroids, Kuiper Belt Objects, small moons),
can hence offer a resolved vertical view of the thermal structure within the
surface.

High spatial resolution mapping of the largest bodies can add the horizontal
component to achieve a full 3-D thermal map of a surface. The objective is then
to link thermal features to insolation effects, geographic features and the
distribution of surface properties (albedo and thermal inertia in particular).
With the ngVLA, the resolution would be sufficient to map Mars, Venus and
Mercury (at all frequencies), large moons and Pluto (at mid and high frequencies),
and main-belt asteroids down to roughly 100 km diameter (at high frequencies).
In addition, the possibility to map the variation of the polarization as a
function of emission angle across a surface is the most direct method to derive
a surface's dielectric constant. 

\bigskip
\textit{Frequency requirement:} Frequency coverage determines the range of depths probed in the subsurface. The possibility of measuring low frequency emission is especially valuable for probing subsurface emission below the diurnal skin depth. 

\bigskip
\textit{Angular resolution requirement:} Resolution down to $\sim$\,50 mas, at least at high frequencies, is a minimum for imaging the surfaces of the largest moons and asteroids. Any improvement over this would allow us to reach for mid-size asteroids.

\bigskip
\textit{Other requirements:} Observations may require stringent scheduling constraints, related to observing geometry.\\

\subsection{Surface and subsurface bistatic radar}

Radar observations have been used to probe the surfaces and subsurfaces of all
of the terrestrial planets and many smaller bodies in the Solar System
\citep{Ostro2003295}.  Notable findings include the first indications of water ice in the
permanently shadowed regions at the poles of Mercury
\citep{1992Sci...258..635S,2001Icar..149....1H}, searches for water ice in the
polar regions of the Moon \citep{1997Sci...276.1527S,2003Natur.426..137C}, and
mapping of polar ice and anomalous surface features on Mars
\citep{1991Sci...253.1508M}.  More recently, there has been considerable effort
in using radar observations both to characterize near-Earth objects (NEOs) and
determine their physical properties, spin, and orbital state
\citep{2002Sci...296.1445M,2010Icar..209..535B}.  Precise
knowledge of their orbits is essential to assess the extent to which they might
represent impact hazards to the Earth \citep{NEO_hazard}, and to support both robotic and crewed spacecraft missions \citep[e.g.,][]{2014Icar..235....5C}.

Radar observations are currently conducted at S~band ($\sim$\,2.3\,GHz)
and~X~band ($\sim$\,8.5\,GHz), and future radar observations may also be
conducted at Ka~band ($\sim$\,32\,GHz).  All of the planetary radar bands could
be within the frequency coverage of the ngVLA. Existing planetary radar
observations with the VLA have been used to image the radar return and make
plane-of-the-sky astrometry measurements.  Such ``bistatic radar aperture
synthesis'' has produced spectacular images that convincingly demonstrate the
presence of water ice at the poles of Mercury and Mars.  With a factor of
\textit{five} better angular resolution, the ngVLA could produce a
correspondingly better linear resolution on the surface of target bodies and
open the possibility of producing resolved images of smaller objects.  For
example, \cite{1994Icar..111..489D} imaged 4179 Toutatis with the VLA with an
angular resolution corresponding to a linear resolution of $\sim$10~km on the
asteroid. The asteroid showed clearly distinct residual radar features,
suggestive of a bi-lobed structure, but the estimate of the separation of these
features was limited by the VLA's beam size.  With the improved angular
resolution of the ngVLA, features with scales of about~2~km would have been
distinguishable.

In addition, ngVLA sensitivity would expand the set of targets for traditional
bistatic delay-Doppler planetary radar.  The signal-to-noise ratio of radar
observations scales as $R^{-4}$, where $R$ is the range to the object.  With a
fixed radar transmitter power (and gain), the signal-to-noise ratio can only be
improved by using increasing the gain (i.e., sensitivity) of the (bistatic)
receiving element. The increased sensitivity of the ngVLA would increase the
range to which NEOs could be targeted for radar observations, particularly for
targets that are outside of the declination range of the Arecibo Observatory. If
the ngVLA reaches a sensitivity of five times that of the current VLA, it would
more than triple the accessible volume (enlarge the range by a factor of~50\%)
for NEO observations. In addition, for NEOs with close approaches to the Earth
(short round-trip light travel times), it can be difficult or impossible to
switch a radar facility from transmitting to receiving rapidly enough, and using
a bistatic configuration is the only way to probe them with radar. 

\bigskip
\textit{Frequency requirement:} S-band ($\sim$\,2.3\,GHz) and X-band ($\sim$\,8.5\,GHz) coverage is desirable.

\bigskip
\textit{Sensitivity requirement:} Sensitivity directly constrains the limit in target distance. The nominal improvement of sensitivity by a factor 5 compared to the JVLA already significantly increases the number of potential targets.

\bigskip
\textit{Angular resolution requirement:} Resolution of the order of 300 mas at S-band ($\sim$\,2.3\,GHz) would allow for high-resolution radar mapping of several near-Earth asteroids.

\bigskip
\textit{Other requirements:} The availability of a bistatic configuration is critical for observing some of the nearest targets.

\subsection{Comets' chemistry and nuclei}

Comets are considered to be among the most primitive objects of the Solar
System. Their rich coma chemistry, which has been explored in depth in
particular at (sub)millimeter wavelengths \citet{1997EM&P...78....5B}, may
contain clues related to the physical and chemical conditions of the primitive
Solar System.  At the ngVLA wavelengths, emission from water, ammonia,
formaldehyde, methanol, OCS, and other coma constituents could be mapped, and
their ratios measured.  In particular, the true ammonia/water ratio could be derived by the simultaneous mapping of the 22~GHz water line and 24~GHz ammonia line in the near-nucleus coma (within 1000 km). This measurement, which is connected to a science case for disk studies with ngVLA, should enable us to explore how comet chemistry is linked to the protoplanetary disk.

If lower frequencies are available, molecules such as OH
and CH could also be mapped, either directly or via the method of occultation
\citep{1997DPS....29.3404B}.  Probing the lower-lying transitions of these
molecules (those not available to ALMA, for instance) is important for
understanding their abundance and temperature.  Unfortunately, these transitions
are fundamentally much weaker than the higher-level transitions, since the lines
are all thermally excited and hence the flux density scales like $\nu^3$.
Because of this, such studies have been very limited in the past(for an
exception, see, e.g., \citealt{1989AJ.....97..246S}).  The high sensitivity
provided by the ngVLA is thus critical for these types of observations.  One
challenge for a number of these molecules is the size of the emitting region,
which can be very large.  In some cases (including ammonia), it may be necessary
to use the ngVLA for the equivalent of single-dish observations, potentially
using position- or frequency-switching.  

Observations of the thermal emission from cometary nuclei is similar in
principle to the observation of surfaces of small bodies, and allows one to
estimate the nucleus' equivalent size and its surface emissivity properties. Due
to the nuclei's icy composition, radio observations usually sound deeper levels
in cometary nuclei than in rocky surfaces. Bistatic radar of comets (see above)
also provides information on the nuclei and their ``icy grain halos''
\citep{2004come.book..265H}.

\bigskip
\textit{Frequency requirement:} K band (18--27\,GHz).

\bigskip
\textit{Sensitivity requirement:} High sensitivity is crucial to detect the ammonia line at a range of heliocentric distances.

\bigskip
\textit{Spectral resolution requirement:} Kinematics studies could be performed on ammonia lines in comets if the lines are sufficiently resolved, requiring a spectral resolution below 0.2\,\kms{} at 24\,GHz. 

\bigskip
\textit{Field of view requirement:} Due to the large extent of comets' comae, mosaicking may be necessary.

\bigskip
\textit{Imaging requirements:} Similar to the giant planet case, simultaneous observations of short and long baselines are a must. Even with short baselines, large spatial scales may still be filtered out, warranting a single-dish observation mode.

\bigskip
\textit{Other requirements:} Observations may require stringent scheduling constraints.

\subsection{Rings}

Much information can be extracted from radio observations of Saturn's rings.
High precision maps of the rings at high spatial resolution and at different
frequencies (2--50~GHz) can be used to determine the mass fraction of non-icy
material in the rings (and how this varies throughout the rings), as well as the
nature and size distribution of the smallest particles in Saturn's rings
\citep{zhang2014}. Similar to atmospheric observations of the giant planets,
observations of Saturn's rings require simultaneous coverage of small and long
baselines. Measuring the polarization of the rings is also useful to determine
particle properties from their scattering characteristics.

\bigskip
\textit{Other requirements:} linear polarization.

\subsection{Jupiter synchrotron emission}

Jupiter's synchrotron emission originates from the interaction of energetic
electrons with Jupiter's strong magnetic field. The synchrotron emission
morphology and intensity is changing over time in response to comet impacts,
interplanetary shocks and other phenomena that  induce changes in the energetic
electron distribution in Jupiter's magnetosphere. With the limited
signal-to-noise in the older VLA maps, and because of changes in Jupiter's size
as the planet orbits the Sun, it has been difficult to assess the reality of
short-term time variations \citep{depater1997}. With the ngVLA sensitivity, such
studies can be carried out. It is ideal to observe over the entire frequency
range, from the lowest frequencies (0.3 GHz) up to $\sim$\,50 GHz. If time
variability is observed, the source and mode of transport of energetic electrons
can be determined---this is one of the outstanding questions in Jovian
magnetospheric physics. Given the large spatial scales involved, the
simultaneous use of extended and short baselines is necessary to study the full
emission extent. 

\bigskip
\textit{Frequency requirement:} Frequency coverage as wide as possible is preferable.

\bigskip
\textit{Sensitivity requirement:} A continuum sensitivity at least 5 times better than JVLA is desirable.

\bigskip
\textit{Field of view requirement:} A large instantaneous field of view ($\sim$1 arcmin) is preferable, and mosaicking will be necessary to map Jupiter's environment.

% SETI
\clearpage
\section{SETI}

SETI (Search for ExtraTerrestrial Intelligence) experiments seek to determine the distribution of advanced life in the universe through detecting the presence of advanced technology, usually by detecting electromagnetic emission from communication technology \citep{Tarter:2003p266}, but also by detecting evidence of large scale energy usage \citep{2009ApJ...698.2075C,2014ApJ...792...27W,2014ApJ...792...26W}, or interstellar propulsion \citep{2015arXiv150803043G}.  For more than 100 years, technology constructed by human beings has been producing radio emission that would be readily detectable at tens of parsecs using receiving technology only moderately more advanced than our own. Some emission, including that produced by the planetary radars at Arecibo Observatory and the NASA Deep Space Network, would be detectable across our galaxy. Modern radio SETI experiments have been ongoing for the last 55 years, but for the most part they have searched only a small fraction of the radio spectrum accessible from the surface of the Earth and have probed only a few nearby stars at high sensitivity. The sensitivity and flexibility of the ngVLA, leveraged by the wealth of new SETI targeting information that will be forthcoming from experiments such as GAIA \citep{2001A&A...369..339P} and the \textit{Transiting Exoplanet Survey Satellite} (TESS) \citep{2014SPIE.9143E..20R}, could enable a vastly more powerful radio SETI search than has been conducted in the past. For example, at 5 $\times$ JVLA sensitivity, ngVLA could detect radio emission of comparable luminosity to our own high-power aircraft radar from dozens of nearby stars in only 10 minutes.\footnote{We note that this statement applies only generally to isotropic transmitters of similar equivalent isotropically radiated power (EIRP), not to the specific case of detecting true Earth-analog radar systems}  Further, the proposed frequency coverage and putative sky coverage of the ngVLA would be very complementary to other upcoming radio SETI facilities, including the SKA \citep{2015aska.confE.116S}.

SETI observations require access to low-level time-domain data products in order to achieve high sensitivity to coherent and/or technologically modulated signals. For interferometers, generally this is accomplished in a manner similar to operating the telescope as a single unified VLBI station---e.g., the signals from all antennas are phased up (or beam-formed) to one or more positions on the sky, and the resultant voltage time series is recorded to disk or analyzed in-situ with one or more custom signal processing systems \citep{2006AcAau..59.1153D}.  Multiple beams on the sky are especially effective because they allow powerful multi-beam interference excision to be employed in the candidate-sorting stage \citep{Barott:2011kt}. Although SETI signal processing systems are custom in the sense that they run specific signal detection algorithms that may differ from those used in other radio astronomy applications, recently the hardware portion has been largely constructed from commodity elements, e.g., rack mount CPUs and GPUs \citep{2014ebi..conf..5.2K}.

One could envision several possible paradigms for SETI observations with the ngVLA, including both commensal and primary-user modes. In the case of the former, SETI observations would be most successful if a capability were built into the overall system that permitted simultaneous beam-forming, and subsequent SETI signal analysis, within the primary beam of the telescope regardless of what other activities were underway. In general, SETI targets are isotropic on the sky.  Although one can make reasonable arguments for concentrating on regions of high galactic stellar density in raster scan surveys, there are sufficient unknowns that a true all-sky survey is at least as effective \citep{2000AcAau..46..649S}. For both commensal and primary-user modes, access to as wide a phased bandwidth as possible will maximize search speed.

\bigskip
\textit{Frequency requirement:}
Large instantaneous bandwidths would allow searching a wide range of possible transmission frequencies quickly.

\bigskip
\textit{Sensitivity requirement:}
As discussed above, at 5 $\times$ JVLA sensitivity, ngVLA could detect radio emission of comparable luminosity to our own high-power aircraft radar from dozens of nearby stars in only 10 minutes.  Higher sensitivity allows probing more Earth-like luminosity functions for artificial transmitters.

\bigskip
\textit{Angular resolution requirement:}
For targeted SETI searches, most sources of interest would be unresolved with any Earth-based radio telescope.  For SETI surveys, it is desirable to have a small number (N $\approx$ 10) of relatively large beams on the sky in order to scan large fields of view quickly without having to search many independent signal paths.

\bigskip
\textit{Spectral resolution requirement:}
The characteristic frequency resolution for narrow-band SETI experiments at cm-wavelengths is $\sim$\,0.1 Hz.

\bigskip
\textit{Field of view requirement:} 
For a targeted SETI search, it would be desirable to have a field of view large enough that there were as many potential SETI targets within the field as there were beams available to search them.  For a catalog of 1M objects distributed isotropically, and 10 available SETI search beams, FoV $\approx$ 0.5 deg$^2$ would be ideal.

% Summary of the technical requirements.
\clearpage
\section{Summary of the technical requirements and conclusions}

We presented a series of science cases that will be made possible by a long-baseline array operating 
at millimeter and centimeter wavelengths, such as the Next Generation Very Large Array (ngVLA).  
Table~\ref{table:technical_requirements} summarizes the technical specifications required to successfully 
achieve the presented scientific goals. The Key Science Projects are listed in bold. 

The majority of the presented  science cases, and in particular all the key science projects, require 
observations between 10--50\,GHz. We note that this range of frequency would be covered only marginally 
by the SKA and the ALMA Band 1 receivers. 

The most compelling science involving the formation of stars and planets is enables by achieving 
a maximum angular resolution between 0.005--0.01\arcsec. This would require baselines 
as long as 400\,km for a nominal frequency of 30\,GHz. Mapping projects, such as mapping planet 
formation regions and the innermost part of dense molecular cores, require high fidelity images, which in turn requires 
good coverage of the uv-plane.  
Deep mapping of the atmosphere of the giant planets in our Solar system, and, in general, Solar 
system related science, requires large field of view capabilities, coupled with intermediate angular 
resolution (0.05--0.5\arcsec).  The request of high angular resolution, large field of view, and good uv-coverage
should be carefully taken into account in designing the array.

Finally, since several of the presented science cases will benefit from complementary  ALMA and SKA observations, 
the location of such a future array should be chosen in order to maximize the overlap of sky coverage. 
%Location of the ngVLA in the southern hemisphere appears therefore to be the most reasonable solution.

\begin{landscape}

\begin{table}
\centerline{Table \ref{table:technical_requirements}: Technical requirements of the ngVLA ``Cradle of Life'' science cases \vspace*{0.1in}}
\begin{center}
{\footnotesize
\begin{tabular}{llccccc}
\hline
(1)			                                                             & (2)       & (3)		      & (4)  		&(5) 			& (6)			& (7) \\ 
Science case 	                                                             &            & Frequency  & Sensitivity & Angular res. & Max. Ang. Scale & Vel. res. \\
                                                                                      &             & (GHz)        & (\ujybm{}) & (mas)     & ($\arcsec$)                       & (km/sec) \\
\hline \\
\S\,2.1 \textbf{HII regions \& high-mass jets}  		& cont	& 10--50     	& 1-10      		& 5--20 	& 5  	& ---  		\\
\S\,2.1 \textbf{HII regions \& high-mass jets}  		& line   	& 40--100 	& 1     	   	& 20          &  5 	  	& 1 		\\
\S\,2.2 Infall via spectral line absorption 	    		& line   	& 18--40   	& 100 K     	& 20         	&  10		&  	 1 	\\
\S\,2.3 Polarization in star-cluster-forming cores  	& cont 	& 40--50  	& ---			& 5--100 	& 5			&	---	\\
\S\,2.4 Star formation in the galactic center  		& cont	& 20--70    & 0.1		& 10--20  	& 2			&  --- 		\\ \\

\S\,3.1 \textbf{Binary/multiple protostars}  			& cont 	& 10--30	& 1--10 K 		& 20		& $<1$ 	&	---	 \\
\S\,3.2 Dust polarization			                   	& cont 	& 20--50  & 1		        & 100--3000     & $<5$         &       ---  	 \\
\S\,3.3 Structure of protostellar jets                            	& cont 	& 4--30  & 0.5 		        & 20--150     & 10         &       ---  	 \\ \\

\S\,4.1 \textbf{Mapping planet formation}  	              	& cont 	& 10-50 	&  1 K	 & 1--10     & $<$5		&	---	\\
\S\,4.2 \textbf{Probing dust/pebble distribution in disks} & cont	& 10--50    &  1 K         & 1--50     & $<$5		&	---	 \\ \\

\S\,5.1 Complex molecules in hot cores         		& line   	& 18--50   	& 0.15 K	& 20--100   	&  $<$5  		&  0.1--0.5 	\\
\S\,5.2 \textbf{Chemistry (NH$_3$) of planet-forming regions} & line 	& 10--40	&  1K		        & 20--5000		& $<$5	    	& 0.1-1	 \\ \\

\S\,6.1 Debris disks 							 & cont      &  10--100 & 0.1               & 1000--5000		&  50		&	--- 	 \\ \\

\S\,7.1 \textbf{Deep atmospheric mapping}  & cont & 1--50 & --- & $<$0.5 & 50 & ---\\
\S\,7.2 (Sub)surface thermal emission  & cont & 10--50 & --- & $<$0.05 & 8 & ---\\
\S\,7.3 (Sub)surface structure using bistatic radar & cont  & 2--8 & --- & $<$0.05 & 1 & --- \\
\S\,7.4 Comet chemistry                     & line   & 22--24 & --- & $<$0.5 & 300 & $<$0.2\\
\S\,7.5 Planetary rings                        & cont  & 10--50 & --- & --- & 120 & ---\\
\S\,7.6 Jupiter synchrotron emission  & cont & 10--50 & --- & --- &  200 & ---\\ \\

\S\,8.1 SETI  								& line 	& 5--100 	& >5$\times$ JVLA & --- & 0.5\,deg$^2$ FOV & 0.1\,Hz \\ \\
\hline
\end{tabular}

\bigskip
\caption{\footnotesize Technical requirements for each of the ngVLA science cases presented by the ``Cradle of Life'' working group.
Key Science Projects (KSPs) are in bold.}
\label{table:technical_requirements}
}
\end{center}
\end{table}

\end{landscape}

% Acks
\section*{Acknowledgments}

The National Radio Astronomy Observatory is a facility of the National Science Foundation operated under cooperative agreement by Associated Universities, Inc.
Part of this research was carried out at the Jet Propulsion Laboratory, California Institute of Technology, under a contract with the National Aeronautics and Space Administration. A. I. acknowledge financial support from the National Science Foundation through the award number AST-1109334  and from NASA through  the award number NNX14AD26G.

\bibliography{ngVLA}
\bibliographystyle{apj}

\end{document}